\documentclass[aps,prx,twocolumn,floatfix,superscriptaddress,amsmath,amssymb,amsfonts]{revtex4-2}

\usepackage{physics}

\usepackage{graphicx}
\usepackage{dcolumn}
\usepackage{amsmath,amsfonts,amssymb}
\usepackage{bm}
\usepackage{xspace}
\usepackage[T1]{fontenc}
\usepackage{newtxtext,newtxmath}
\usepackage{xcolor}
\usepackage{dsfont}
\usepackage{upgreek}
\usepackage{orcidlink} 

\usepackage{tabularx}
\usepackage{multirow, makecell}
\usepackage{array}
\usepackage{stackengine}

\setcitestyle{super}

\makeatletter
\renewcommand\@make@capt@title[2]{%
	\@ifx@empty\float@link{\@firstofone}{\expandafter\href\expandafter{\float@link}}%
	{\textbf{#1}}\@caption@fignum@sep#2\quad
}%
\makeatother

\makeatletter 
\renewcommand{\fnum@figure}{\textbf{Fig.~\thefigure}}
\renewcommand*{\@caption@fignum@sep}{\textbf{ | }}
\makeatother

\renewcommand{\thetable}{\arabic{table}}
\makeatletter 
\renewcommand{\fnum@table}{\textbf{Table~\thetable}}
\makeatother

\definecolor{Aqua}{RGB}{15,15,180}
\hypersetup{hidelinks}

\newcommand{\Tc}{\ensuremath{T_{\mathrm{c}}}\xspace}

\newcommand{\lamL}{\ensuremath{\lambda_{\mathrm{L}}}\xspace}

\newcommand{\jdp}{\ensuremath{j_{\mathrm{dp}}}\xspace}

\newcommand{\ACchem}{A$_3$C$_{60}$\xspace}

\begin{document}
	
	\title{Bypassing the lattice BCS--BEC crossover in strongly correlated superconductors through multiorbital physics}
	
	\author{Niklas Witt\,\orcidlink{0000-0002-2607-4986}}
	\email{niklas.witt@physik.uni-hamburg.de}
	\affiliation{I. Institute of Theoretical Physics, University of Hamburg, Notkestraße 9-11, 22607 Hamburg, Germany}
	\affiliation{The Hamburg Centre for Ultrafast Imaging, Luruper Chaussee 149, 22607 Hamburg, Germany}
	
	\author{Yusuke Nomura\,\orcidlink{0000-0003-4956-562X}}
	\affiliation{Institute for Materials Research (IMR), Tohoku University, 2-1-1 Katahira, Aoba-ku, Sendai 980-8577, Japan}
	
	\author{Sergey Brener}
	\affiliation{I. Institute of Theoretical Physics, University of Hamburg, Notkestraße 9-11, 22607 Hamburg, Germany}
	
	\author{Ryotaro Arita\,\orcidlink{0000-0001-5725-072X}}
	\affiliation{Department of Physics, The University of Tokyo, 7-3-1 Hongo, Bunkyo-ku, Tokyo, 113-0033, Japan}
	\affiliation{RIKEN Center for Emergent Matter Science (CEMS), 2-1 Hirosawa, Wako, Saitama 351-0198, Japan}
	
	\author{Alexander I. Lichtenstein\,\orcidlink{0000-0003-0152-7122}}
	\affiliation{I. Institute of Theoretical Physics, University of Hamburg, Notkestraße 9-11, 22607 Hamburg, Germany}
	\affiliation{The Hamburg Centre for Ultrafast Imaging, Luruper Chaussee 149, 22607 Hamburg, Germany}
	
	\author{Tim O. Wehling\,\orcidlink{0000-0002-5579-2231}}
	\affiliation{I. Institute of Theoretical Physics, University of Hamburg, Notkestraße 9-11, 22607 Hamburg, Germany}
	\affiliation{The Hamburg Centre for Ultrafast Imaging, Luruper Chaussee 149, 22607 Hamburg, Germany}
	
	\date{\today}
	
	
	\begin{abstract}
		Superconductivity emerges from the spatial coherence of a macroscopic condensate of Cooper pairs. Increasingly strong binding and localization of electrons into these pairs compromises the condensate's phase stiffness, thereby limiting critical temperatures -- a phenomenon known as the BCS--BEC crossover in lattice systems. In this study, we demonstrate enhanced superconductivity in a multiorbital model of alkali-doped fullerides (\ACchem) that goes beyond the limits of the lattice BCS--BEC crossover. We identify that the interplay of strong correlations and multiorbital effects results in a localized superconducting state characterized by a short coherence length but robust stiffness and a domeless rise in critical temperature with increasing pairing interaction. To derive these insights, we introduce a new theoretical framework allowing us to calculate the fundamental length scales of superconductors, namely the coherence length ($\xi_0$) and the London penetration depth (\lamL), even in presence of strong electron correlations.
	\end{abstract}
	
	\maketitle

	\section*{Introduction}
	The collective and phase coherent condensation of electrons into bound Cooper pairs leads to the emergence of superconductivity. This macroscopic coherence enables dissipationless charge currents, perfect diamagnetism, fluxoid quantization and technical applications~\cite{Tokura2017,Yao2021} ranging from electromagnets in particle accelerators to quantum computing hardware. Often, superconducting (SC) functionality is controlled by the critical surface spanned by critical magnetic fields, currents, and temperatures which a SC condensate can tolerate. Fundamentally, these are determined by the characteristic length scales of a superconductor -- the London penetration depth, \lamL, and the coherence length, $\xi_0$.
	
	\lamL and $\xi_0$ quantify different aspects of the SC condensate: The penetration depth is the length associated with the mass term that the vector potential gains through the Anderson-Higgs mechanism~\cite{Shimano2020}. In consequence, magnetic fields decay exponentially over a distance \lamL inside a superconductor. Through this, \lamL is connected to the energy cost of order parameter (OP) phase variations and hence the SC stiffness $D_{\mathrm{s}}$. The coherence length, on the other hand, is the intrinsic length scale of OP amplitude variations and is associated with the amplitude Higgs mode. $\xi_0$ sets the scale below which amplitude and phase modes significantly couple such that spatial variations of the OP’s phase reduce its amplitude~\cite{Shimano2020,Coleman2008}.
	
\begin{figure*}[t]
		\includegraphics[clip,width=1\textwidth]{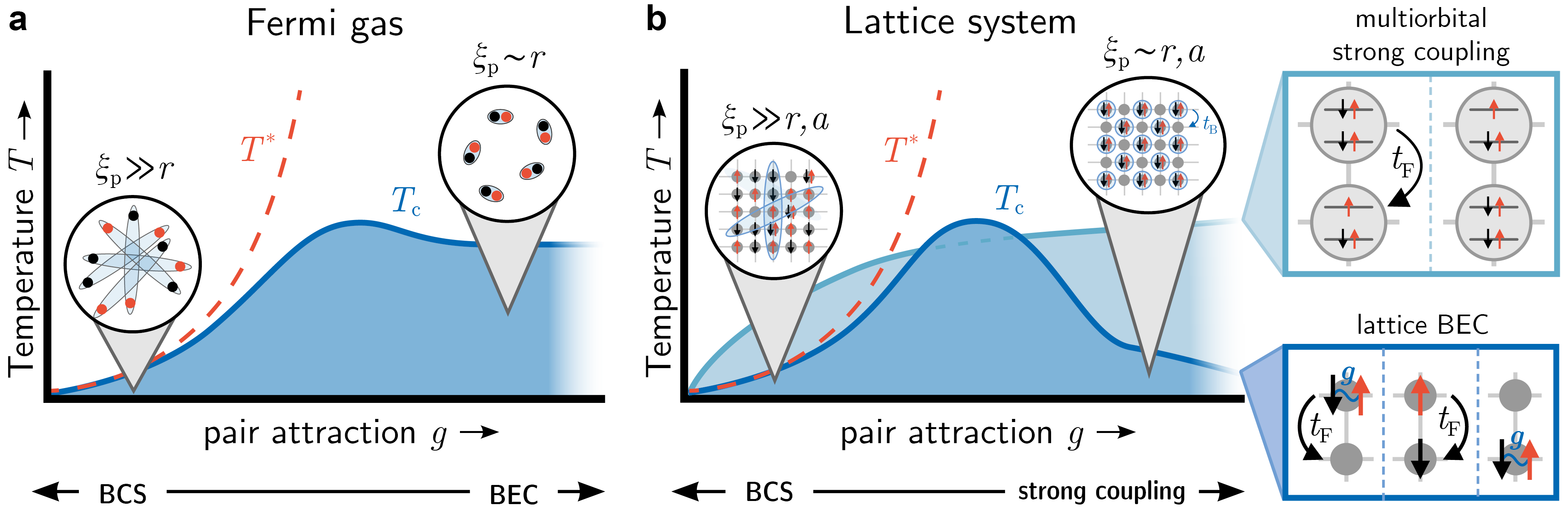}
		\caption{\textbf{BCS--BEC crossover in Fermi gases vs. lattice systems.} Evolution of the critical temperature \Tc and Cooper pair size $\xi_{\mathrm{p}}$ in the BCS--BEC crossover (dark blue line) for (\textbf{a}) Fermi gases and (\textbf{b}) lattice systems. Both cases display a dome-shaped behavior of \Tc in the intermediate crossover regime but behave qualitatively different in the strong coupling BEC phase: \Tc remains finite in the Fermi gas (\textbf{a})~\cite{SadeMelo1993} but approaches zero in the lattice case (\textbf{b}). During the crossover, $\xi_{\mathrm{p}}$ is reduced to the order of interparticle spacing  $r$ (\textbf{a \& b}) and lattice constant $a$ (\textbf{b}). For the lattice system, we contrast the evolution towards the multiorbital strong coupling phase (light blue line) discussed in this article. Here, the localization of pairs differently affects the bosonic hopping $t_{\mathrm{B}}$, as drawn in the insets on the right. In the lattice BEC limit, bosonic hopping relies on a second-order process involving two fermionic hoppings, $t_{\mathrm{F}}$, and a virtual intermediate state with broken pairs that is inhibited by the strong attraction $g < 0$. Consequently, this process and also $\Tc\propto t_{\mathrm{B}}=t_{\mathrm{F}}^2/|g|$ are quenched at large $|g|$~\cite{Chen2024a,Nozieres1985}. In the multiorbital strong coupling case, the local coexistence of paired and unpaired electrons fluctuating between different orbitals enables bosonic hopping $t_{\mathrm{B}}$ as a first order process in $t_{\mathrm{F}}$ without any intermediate broken pair states. See Supplementary Note 7 for details on the involved states and energy diagram specific to the case of A$_3$C$_{60}$. A second temperature scale $T^*$ is drawn in both panels as its splitting from \Tc (corresponding to the opening of a pseudogap) marks the beginning of the crossover regime~\cite{Chen2024a}.}
		\label{Fig1} 
	\end{figure*}

	In addition to influencing the macroscopic properties of superconductors, \lamL and $\xi_0$ play an important role to understand strongly correlated superconductors, as is epitomized in the Uemura plot~\cite{Uemura1989,Uemura1991,Uemura1991a,Uemura2019}. For instance, the interplay of \lamL and $\xi_0$ impacts critical temperatures~\cite{Emery1995}, it is relevant for the pseudogap formation~\cite{Timusk1999,Norman2005,Lee2006,Gunnarsson2015}, it influences magneto-thermal transport properties like the Nernst effect~\cite{Uemura2019,Ussishkin2002,Jotzu2023}, and it might underlie the light-enhancement of superconductivity~\cite{Jotzu2023,Fausti2011,Mitrano2016,Rowe2023,Cavalleri2017}. An important concept in this context is the BCS--BEC crossover phenomenology~\cite{Leggett1980,Nozieres1985,SadeMelo2008,Randeria2014,Chen2024a}. It continuously connects the two limiting cases of weak-coupling Bardeen--Cooper--Schrieffer (BCS) superconductivity with weakly-bound and largely overlapping Cooper pairs to tightly-bound molecule-like pairs in the strong-coupling Bose--Einstein condensate (BEC) as the interaction strength or the density is varied (Fig.~\ref{Fig1}). 
	
	The BCS--BEC crossover has been studied in ultracold Fermi gases~\cite{Randeria2014}, low-density doped semiconductors~\cite{Nakagawa2021}, and is under debate for several unconventional superconductors~\cite{Uemura1991,Uemura2019,Chen2024a,Park2021,Suzuki2022,Sous2023,Chen2024,Mizukami2023}. However, quasi-continuous systems, for instance Fermi gases, and strongly correlated superconducting solids show a crucially different behavior of how their SC properties, most importantly the critical temperature \Tc, change towards the strong coupling BEC limit. While \Tc converges to a constant temperature $T_{\mathrm{BEC}}$ for Fermi gases in a continuum~\cite{SadeMelo1993} (Fig.~\ref{Fig1}a), \Tc can become arbitrarily small in strongly correlated lattice systems due to the quenching of kinetic energy (Fig.~\ref{Fig1}b). Since the movement of electron pairs, i.e., bosonic hopping, necessitates intermediate fermionic hopping, it becomes increasingly unfavorable for strong attractions. Thus, as Cooper pairs become localized on the scale of the lattice constant, the condensate’s stiffness and hence \Tc are compromised~\cite{Chen2024a,Nozieres1985}. Fig.~\ref{Fig1} contrasts this generic BCS--BEC crossover picture for Fermi gases and correlated lattice systems in terms of the change of \Tc and the pair size $\xi_{\mathrm{p}}$ as a function of pairing strength. Due to the decrease of \Tc in the BCS and BEC limits, a prominent dome-shape of \Tc can be expected in the crossover region for solid materials. Because of this, recent experimental efforts to increase \Tc concentrate on stabilizing SC materials in this region~\cite{Chen2024a,Nakagawa2021,Suzuki2022,Mizukami2023}.

	In this work, we demonstrate how multiorbital effects\cite{Iskin2006,Iskin2011} can enhance superconductivity beyond the expectations of the lattice BCS--BEC crossover phenomenology (as contrasted in Fig.~\ref{Fig1}b) with a model inspired by alkali-doped fullerides (\ACchem with A=K, Rb, Cs). The material family of \ACchem hosts exotic $s$-wave superconductivity of critical temperatures up to $\Tc = 38$\,K, being the highest temperatures among molecular superconductors~\cite{Hebard1991,Ganin2010,Zadik2015,Ramirez2015}, and they possibly reach photo-induced SC at even higher temperatures~\cite{Jotzu2023,Mitrano2016,Rowe2023,Budden2021}. In order to theoretically characterize the SC state, the knowledge of the intrinsic SC length scales is essential. While BCS theory and Eliashberg theory provide a microscopic description of \lamL and $\xi_0$ for weakly correlated materials~\cite{Bardeen1957,Tinkham2004,Allen2008}, their validity is unclear for superconductors with strong electron correlations. To the best of our knowledge, $\xi_0$ is generally not known from theory in strongly correlated materials. Only approaches to determine \lamL exist where an approximate, microscopic assessment of the SC stiffness from locally exact theories has been established, albeit neglecting vertex corrections~\cite{Toschi2005,Yue2021,Gull2013,Simard2019,Harland2019,Liang2017}.
	
	For this reason, we introduce a novel theoretical framework to microscopically access \lamL and $\xi_0$ from the tolerance of SC pairing to spatial OP variations. Central to our approach are calculations in the superconducting state under a constraint of finite-momentum pairing (FMP). Via Nambu--Gor’kov Green functions we get direct access to the superconducting OP and the depairing current \jdp which in turn yield $\xi_0$ and \lamL. The FMP constraint is the SC analog to planar spin spirals applied to magnetic systems~\cite{Liechtenstein1984,Fleck1998,Sandratskii1998}. As in magnetism, a generalized Bloch theorem holds that allows us to consider FMP without supercells; see Supplementary Note 2 for a proof. As a result, our approach can be easily embedded in microscopic theories and \textit{ab initio} approaches to tackle material-realistic calculations.
	
	In this work, we implement FMP in Dynamical Mean-Field theory (DMFT)~\cite{Georges1996} to treat strongly correlated superconductivity. In DMFT, the interacting many-body problem is solved by self-consistently mapping the lattice model onto a local impurity problem. By this, local correlations are treated exactly. We apply the FMP-constrained DMFT to \ACchem where we find $\xi_0$ and \lamL in line with experiment for model parameter ranges derived from \textit{ab initio} estimates, validating our approach. For enhanced pairing interaction, we then reveal a multiorbital strong coupling SC state with minimal $\xi_0$ on the order of only 2\,--\,3 lattice constants, but with robust stiffness $D_{\mathrm{s}}$ and high \Tc which increases with the pair interaction strength without a dome shape. This strong coupling SC state is distinct to the lattice BCS--BEC crossover phenomenology showing promising routes to optimize superconductors with higher critical temperatures. We discuss this possibility in-depth after the presentation of our results.
	
	The remaining paper is organized as follows: First, we motivate how the FMP constraint is linked to \lamL and $\xi_0$ from phenomenological Ginzburg--Landau theory after which we summarize the microscopic approach from Green function methods. Technical details are available in the Supplementary Information as cross-referenced at relevant points. Subsequently, we discuss our results for the multiorbital model of \ACchem. Readers primarily interested in the analysis of these results might skip the first part on the physical motivation for obtaining superconducting length scales from the FMP constraint.
	

	\section*{Results}
	\subsection*{Extracting superconducting length scales from the constraint of finite-momentum pairing}
	
\begin{figure}[t]
		\centering
		\includegraphics[clip,width=1\columnwidth]{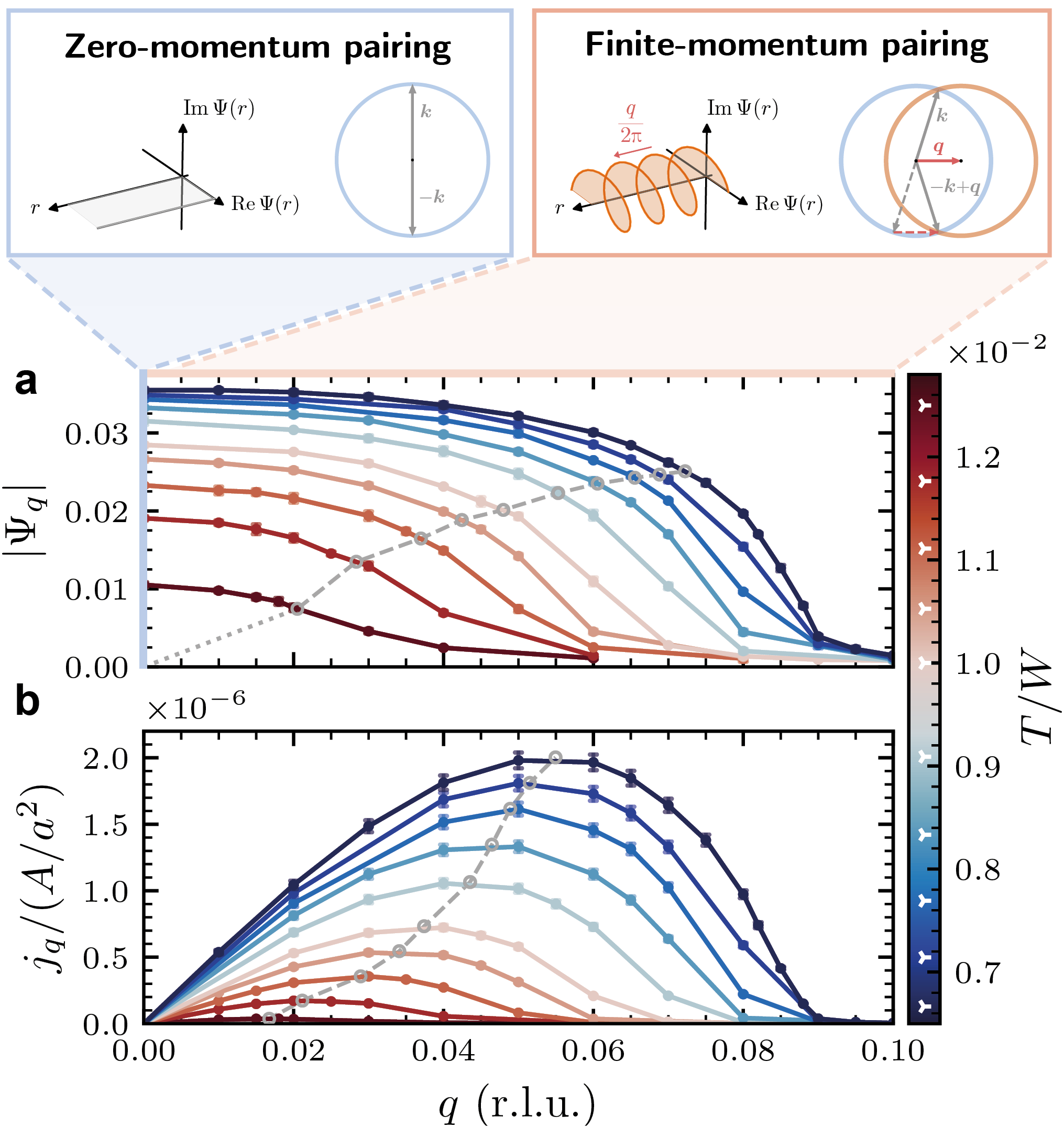}
		\caption{\textbf{Influence of finite-momentum pairing (FMP) constraint on the superconducting condensate.} The top panel insets sketch the position and momentum space representation of the order parameter (OP) $\Psi_{\bm{q}}(\bm{r})=|\Psi_{\bm{q}}|\mathrm{e}^{i\bm{qr}}$ in the zero-momentum (left, $q=0$) and finite-momentum pairing states (right, $q > 0$). The main panels show (\textbf{a}) the momentum dependence of the OP modulus and (\textbf{b}) the supercurrent density $j_q=|\bm{j}_{\bm{q}}|$ as function of Cooper pair momentum $q=|\bm{q}|$ in reciprocal lattice units (r.l.u.). Gray lines indicate the points of extracting $\xi$ and \jdp (see text). The data shown are results for the \ACchem model (c.f.~Eq.~(\ref{eq:KH_interaction})) with lattice constant $a$ and interaction parameters $U/W=1.4$, $J/W=-0.04$ evaluated at different temperatures $T$ (color coded; see color bar with white triangular markers).}
		\label{Fig2} 
	\end{figure}

	In most SC materials, Cooper pairs do not carry a finite center-of-mass momentum $\bm{q}=\bm{0}$. Yet, in presence of external fields, coexisting magnetism, or even spontaneously SC states with FMP, i.e., $\bm{q}\neq\bm{0}$, might arise~\cite{Chen2018,Zhao2023,Wan2023,Yuan2022,Zhang2024} as originally conjectured in Fulde--Ferrel--Larkin--Ovchinnikov (FFLO) theory~\cite{Fulde1964,Larkin1964,Kinnunen2018}. 
	Here, we introduce a method to access the characteristic SC length scales $\xi_0$ and \lamL of strongly correlated materials through the calculation of FMP states with a manually constrained pair center-of-mass momentum $\bm{q}$\cite{Tinkham2004,Coleman2008,Abrikosov1975,Bardeen1962,Fisher1973}. For this, we enforce the OP to be of the FMP form $\Psi_{\bm{q}}(\bm{r})=|\Psi_{\bm{q}}|\mathrm{e}^{i\bm{qr}}$ corresponding to FF-type pairing~\cite{Fulde1964}, which is to be differentiated from pair density waves with amplitude modulations~\cite{Larkin1964,Kinnunen2018,Agterberg2020}. We contrast the OP for zero and finite momentum in real and momentum space in the top panel of Fig.~\ref{Fig2}. For FMP, the OP’s phase is a helix winding around the direction of $\bm{q}$, while the OP for zero-momentum pairing is simply a constant.
	
	Before turning to the implementation in microscopic approaches, we motivate how the FMP constraint relates to \lamL and $\xi_0$. The Ginzburg--Landau (GL) framework provides an intuitive picture to this connection\cite{Tinkham2004,Coleman2008,Bardeen1962} which we summarize here and discuss in detail in Supplementary Note 1, including a derivation of the following equations. The GL low-order expansion of the free energy density $f_{\mathrm{GL}}$ in terms of the FMP-constrained OP reads
	\begin{align}
		f_{\mathrm{GL}}[\Psi_{\bm{q}}] = \alpha|\Psi_{\bm{q}}|^2 + \frac{b}{2}|\Psi_{\bm{q}}|^4 + \frac{\hbar^2}{2m^*} q^2 |\Psi_{\bm{q}}|^2
		\label{eq:GL_free_energy}
	\end{align}
	with $q = |\bm{q}|$. $\alpha$, $b$, and $m^*$ are the material and temperature dependent GL parameters. Here, the temperature-dependent (GL) correlation length $\xi(T)$ appears as the natural length scale of the amplitude mode ($\propto |\alpha|$) and the kinetic energy term
	\begin{align}
		\xi(T) = \sqrt{\frac{\hbar^2}{2m^*|\alpha|}} = \xi_0 \left(1-\frac{T}{T_{\mathrm{c}}}\right)^{-\frac{1}{2}}
		\label{eq:coherence_length}
	\end{align}
	with its zero-temperature value $\xi_0$ being the coherence length~\cite{Coleman2008}. The stationary point of Eq.~(\ref{eq:GL_free_energy}) shows that the $\bm{q}$-dependent OP amplitude $\left|\Psi_{\bm{q}}\right|=\left|\Psi_0\right|\sqrt{1-\xi^2q^2}$ ($T$-dependence suppressed) decreases with increasing momentum $q$. For large enough $q$, SC order breaks down (i.e., $\Psi_{\bm{q}}\to0$) as the kinetic energy from phase modulations becomes comparable to the gain in energy from pairing. The length scale associated with this breakdown is $\xi(T)$ and can, therefore, be inferred from the $q$-dependent OP suppression. We employ, here, the criterion $\xi(T)=1/(\sqrt{2}|\bm{Q}|)$ with $\bm{Q}$ such that $|\Psi_{\bm{Q}}(T)/\Psi_0(T)|=1/\sqrt{2}$ for fixed $T$ (see Supplementary Note 5-A for more information).
	
	The finite center-of-mass momentum of the Cooper pairs entails a charge supercurrent $\bm{j}_{\bm{q}}\propto |\Psi_{\bm{q}}|^2\bm{q}$, c.f.~Eq.~(8) in Supplementary Note 1. This current density is a non-monotonous function of $q$ with a maximum called depairing current density \jdp. It provides a theoretical upper bound to critical current densities, $j_{\mathrm{c}}$, measured in experiment. We note that careful design of SC samples is necessary for $j_{\mathrm{c}}$ reaching \jdp as its value crucially depends on sample geometry and defect densities~\cite{Tinkham2004,Bardeen1962,Xu2010}. As the current is related to the London penetration depth \lamL through the second London equation $\mu_0\bm{j} = - \lambda_{\mathrm{L}}^{-2}\bm{A}$, we can determine \lamL from \jdp in GL theory via
	\begin{align}
		\lambda_{\mathrm{L}}(T) = \sqrt{\frac{\Phi_0}{3\sqrt{3}\pi\mu_0\xi(T) j_{\mathrm{dp}}(T)}} = \lambda_{\mathrm{L},0}\left(1-\left(\frac{T}{T_{\mathrm{c}}}\right)^4\right)^{-\frac{1}{2}}
		\label{eq:penetration_depth}
	\end{align}
	with the magnetic flux quantum $\Phi_0 = h/2e$. The temperature dependence with the quartic power stated here is empirical~\cite{Bardeen1957,Tinkham2004} and we find that it describes our calculations better than the linearized GL expectation as discussed in Supplementary Note 5-B. Note that taking the $q\to0$ limit in our approach is related to linear-response-based methods to calculate \lamL or, equivalently, the stiffness $D_{\mathrm{s}}$~\cite{Fisher1973,Liang2017}.

	The GL analysis shows that the OP suppression and supercurrent induced by the FMP constraint connect to $\xi_0$ and \lamL. In a microscopic description, we acquire the OP and supercurrent density from the Nambu--Gor’kov Green function
	\begin{align}
		\mathcal{G}_{\bm{q}}(\tau,\bm{k}) &= -\langle T_{\tau}\psi^{}_{\bm{k},\bm{q}}(\tau)\psi^{\dagger}_{\bm{k},\bm{q}}\rangle \notag\\
		&= \begin{pmatrix}
			G_{\bm{q}}(\tau,\bm{k})   &F_{\bm{q}}(\tau,\bm{k})\\
			F^{*}_{\bm{q}}(\tau,\bm{k}) &\bar{G}_{\bm{q}}(\tau,-\bm{k})
		\end{pmatrix}
		\label{eq:NG_GF}
	\end{align}
	where $\psi_{\bm{k},\bm{q}}^\dag=\left(c_{\bm{k}+\frac{\bm{q}}{2}\uparrow}^\dag\ \ c_{-\bm{k}+\frac{\bm{q}}{2}\downarrow}\right)$ (orbital indices suppressed) are Nambu spinors that carry an additional dependence on $\bm{q}$ due to the FMP constraint. $G$ ($F$) denotes the normal (anomalous) Green function component for electrons ($G$) and holes ($\bar{G}$) in imaginary time $\tau$. For $s$-wave superconductivity as in \ACchem~\cite{Capone2002,Capone2009,Nomura2015a,Nomura2016}, we use the orbital-diagonal, local anomalous Green function as the OP
	\begin{align}
		|\Psi_{\bm{q}}| \equiv [F^{\mathrm{loc}}_{\bm{q}}(\tau=0^-)]_{\alpha\alpha} = \sum_{\bm{k}} \langle c_{\alpha\bm{k}+\frac{\bm{q}}{2}\uparrow} c_{\alpha-\bm{k}+\frac{\bm{q}}{2}\downarrow} \rangle\;,
	\end{align}
	which is the same for all orbitals $\alpha$. This allows us to work with a single-component OP. The current density can be calculated via (c.f.~Eq.~(26) in the Supplementary Information)
	\begin{align}
		\bm{j}_{\bm{q}} = \frac{2e}{N_{\bm{k}}} \sum_{\bm{k}} \mathrm{Tr}_{\alpha}\left[\underline{\bm{v}}(\bm{k}) \underline{G}_{\bm{q}}\left(\tau=0^-,\bm{k}-\frac{\bm{q}}{2}\right)\right]
		\label{eq:current_from_GF}
	\end{align}
	where $\hbar\underline{\bm{v}} = \nabla_{\bm{k}}\underline{h}(\bm{k})$ is the group velocity obtained from the one-electron Hamiltonian $\underline{h}(\bm{k})$ and the trace runs over the orbital indices of $\underline{\bm{v}}$ and $\underline{G}_{\bm{q}}$. Underlined quantities indicate matrices in orbital space, $N_{\bm{k}}$ is the number of momentum points, and $e$ is the elementary charge. See Methods and Supplementary Note 4 for details on the DMFT-based implementation and Supplementary Notes 3 for a derivation and discussion of Eq.~(\ref{eq:current_from_GF}).

	The bottom panels of Fig.~\ref{Fig2} show an example of our DMFT calculations which illustrates the $\bm{q}$-dependence of the OP amplitude and current density for different temperatures $T$. Throughout the paper, we choose $\bm{q}$ parallel to a reciprocal lattice vector $\bm{q} = q\bm{b}_1$. We find a monotonous suppression of the OP with increasing $q$. The supercurrent initially grows linearly with $q$, reaches its maximum \jdp and then collapses upon further increase of $q$. Thus, both $|\Psi_q|$ and $j_q$ behave qualitatively as expected from the GL description. For decreasing temperature, the point where the OP gets significantly suppressed moves towards larger momenta $q$ (smaller length scales $\xi(T)$), while $\jdp(T)$ increases. We indicate the points where we extract $\xi(T)$ and $\jdp(T)$ with gray circles connected by dashed lines.

	
	\subsection*{Superconducting coherence in alkali-doped fullerides}
	We apply FMP superconductivity to study a degenerate three-orbital model $H = H_{\mathrm{kin}} + H_{\mathrm{int}}$, where $H_{\mathrm{kin}}$ is the kinetic energy and the electron-electron interaction is described by a local Kanamori--Hubbard interaction~\cite{Georges2013}
	\begin{align}
		H_{\mathrm{int}} = (U-3J)\frac{\hat{N}(\hat{N}-1)}{2} - J\left( 2\hat{\bm{S}}^2 + \frac{1}{2}\hat{\bm{L}}^2 - \frac{5}{2}\hat{N}\right)
		\label{eq:KH_interaction}
	\end{align}
	with total number $\hat{N}$, spin $\hat{\bm{S}}$, and angular momentum operator $\hat{\bm{L}}$. The independent interactions are the intraorbital Hubbard term $U$ and Hund exchange $J$, as we use the SU(2)$\times$SO(3) symmetric parametrization. This model is often discussed in the context of Hund’s metal physics relevant to, e.g., transition-metal oxides like ruthenates with partially filled $t_{2g}$ shells~\cite{Georges2013,Medici2011}. In the special case of negative exchange energy $J < 0$, it has been introduced to explain superconductivity in \ACchem materials~\cite{Capone2002,Capone2009,Nomura2015a,Nomura2016}. In fullerides, exotic $s$-wave superconductivity exists in proximity to a Mott-insulating (MI)~\cite{Ganin2010,Zadik2015,Ihara2010} and a Jahn--Teller metallic phase~\cite{Zadik2015,Suzuki2000,Hoshino2017,Hoshino2019}. The influence of strong correlation effects and inverted Hund’s coupling were shown to be essential for the SC pairing~\cite{Capone2002,Capone2009,Nomura2015a,Nomura2016,Capone2001} utilizing orbital fluctuations~\cite{Koga2015} in a Suhl--Kondo mechanism~\cite{Suzuki2000}. In the weakly correlated regime of $U\to0$ and small $J$, the system follows BCS phenomenology~\cite{Capone2009}.
	
	The inversion of $J$ seems unusual from the standpoint of atomic physics, where it dictates the filling of atomic shells via Hund’s rules. In \ACchem, a negative $J$ is induced by the electronic system coupling to intramolecular Jahn--Teller phonon modes~\cite{Capone2001,Han2003,Nomura2015b,Nomura2016}. As a result, Hund’s rules are inverted such that states which minimize first spin $\bm{S}$ and then angular momentum $\bm{L}$ are energetically most favorable, see the second term of Eq.~(\ref{eq:KH_interaction}) for $J < 0$.
	
\begin{figure}[t]
		\centering
		\includegraphics[clip,width=1\columnwidth]{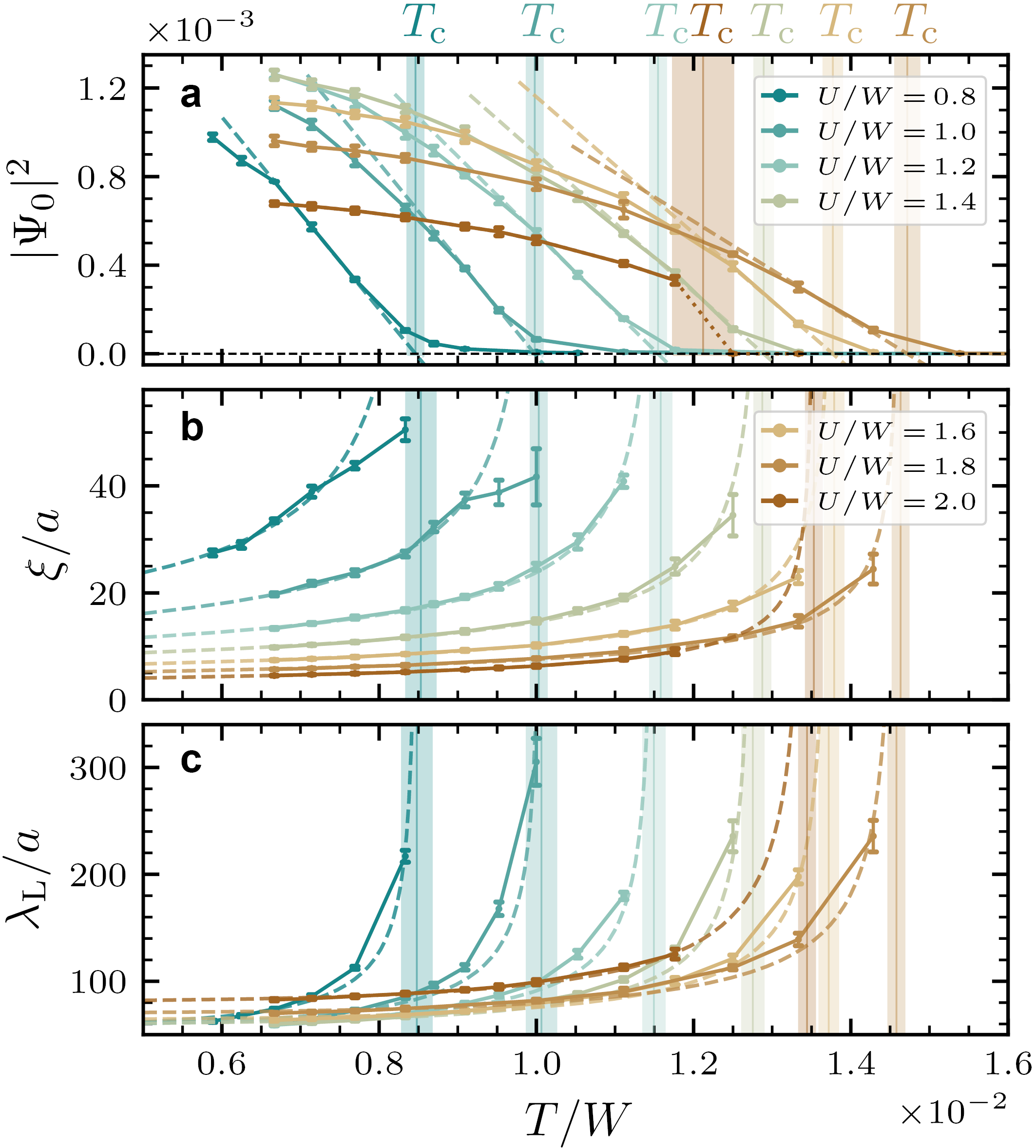}
		\caption{\textbf{Order parameter, correlation length, and penetration depth in \ACchem.} The temperature dependence of (\textbf{a}) the zero-momentum order parameter $|\Psi_0|$, (\textbf{b}) the correlation length $\xi$, and (\textbf{c}) the London penetration depth \lamL are shown for different ratios $U/W$. The data were obtained for fixed $J/W = -0.04$ as estimated from \textit{ab initio} data. Length scales are given in units of the lattice constant $a$. Fits to extract critical temperatures \Tc and zero-temperature values $\xi_0$ and $\lambda_{\mathrm{L},0}$ are shown by dashed lines. The region of uncertainty to fit \Tc is indicated by shaded regions.}
		\label{Fig3} 
	\end{figure}

	We connect the model to \ACchem by using an \textit{ab initio} derived model for the kinetic energy~\cite{Nomura2012}
	\begin{align}
		\begin{split}
			H_{\mathrm{kin}} &= \sum_{ij}\sum_{\alpha\gamma\sigma} t_{\alpha\gamma}(\bm{R}_{ij}) c^{\dagger}_{i\alpha\sigma}c^{}_{j\gamma\sigma}
			\label{eq:kinetic_energy}
		\end{split}
	\end{align}
	where $t_{\alpha\gamma}(\bm{R}_{ij})$ is the hopping amplitude between half-filled $t_{1u}$ orbitals ($\alpha,\gamma=1,2,3$) of C$_{60}$ molecules on sites connected by lattice vector $\bm{R}_{ij}$. We take the bandwidth $W$ as the unit of energy ($W\approx 0.3$\,--\,$0.5$\,eV for Cs to K based \ACchem)~\cite{Nomura2016,Nomura2012}, see Supplemental Note 4-D for further details. For the interaction, we take the first principles’ estimates of fixed $J/W = -0.04$ and varying $U/W$~\cite{Capone2002,Capone2009,Nomura2015a,Nomura2016,Hoshino2017} to emulate unit cell volumes as resulting from the size of different alkali dopants. We solve this Hamiltonian using DMFT, which explicitly takes into account superconducting order. Through the momentum dependence of $|\Psi_{\bm{q}}(T)|$ and ${\bm{j}}_{\bm{q}}(T)$, $\xi(T)$ and $\lamL(T)$ can be extracted as discussed in the previous subsection. We show the derived temperature dependence of $|\Psi_{q=0}(T)|$, $\xi(T)$, and $\lamL(T)$ for different $U/W$ in Fig.~\ref{Fig3}. 
	
	Close to the transition point, the OP vanishes and the critical temperature \Tc can be extracted from  $|\Psi_0(T)|^2 \propto T-\Tc$. We find that \Tc increases with $U$, contrary to the expectation that a repulsive interaction should be detrimental to electron pairing. This behavior is well understood in the picture of strongly correlated superconductivity: As the correlations quench the mobility of carriers, the effective pairing interaction $\sim J/ZW$ increases due to a reduction of the quasiparticle weight $Z$~\cite{Capone2002,Capone2009,Nomura2015a}. The trend of increasing \Tc is broken by a first-order SC to MI phase transition for critical $U\sim2$\,$W$ which is indicated by a dotted line. Upon approaching the MI phase, the magnitude of $|\Psi_0|$ behaves in a dome-like shape. The \Tc values of 0.8\,--\,$1.4 \times 10^{-2}$\,$W$ that we obtain from DMFT correspond to 49\,--\,85\,K which is on the order of but quantitatively higher than the experimentally observed values. The reason for this is that we approximate the interaction to be instantaneous as well as that we neglect disorder effects and non-local correlations which reduce \Tc~\cite{Nomura2015a,Nomura2016}. The same effects could mitigate the dominance of the MI phase which prevents us from observing the experimental \Tc-dome~\cite{Zadik2015,Nomura2016}.
	
	Turning to the correlation length, we observe that, away from \Tc, $\xi(T)$ is strongly reduced to only a few lattice constants ($a\sim14.2$\,--\,14.5\,\AA) by increasing $U$, i.e., pairing becomes very localized. At the same time, $\lambda_{\mathrm{L}}(T)$ is enlarged. Hence, the condensate becomes much softer as there is a reduction of the SC stiffness $D_{\mathrm{s}}\propto\lamL^{-2}$ upon increasing $U$. Fitting Eqs.~(\ref{eq:coherence_length}) and (\ref{eq:penetration_depth}) to our data, we obtain zero-temperature values of $\xi_0=3$\,--\,10\,nm and $\lambda_{\mathrm{L},0}=80$\,--\,120\,nm. Comparing our results with experimental values of $\xi_0\sim2$\,--\,4.5\,nm and $\lambda_{\mathrm{L}}\sim 200$\,--\,800\,nm~\cite{Uemura1991,Uemura1991a,Ramirez2015,Kasahara2017,Gunnarsson1997},  we see an almost quantitative match for $\xi_0$ and a qualitative match for \lamL. Both experiment and theory consistently classify \ACchem as type II superconductors ($\lamL \gg \xi_0$)~\cite{Gunnarsson1997,Ramirez2015}. 
	
	We speculate that disorder and spontaneous orbital-symmetry breaking~\cite{Hoshino2019} in the vicinity of the Mott state could lead to a further reduction of $\xi_0$ as well as an increase of $\lamL$ beyond what is found here for the pure system. This could bring our calculations with minimal $\xi_0=3$\,nm closer to the experimental minimal coherence length of 2\,nm revealed by measurements of large upper critical fields reaching up to a maximal $H_{\mathrm{c}2}=90$\,T in Ref.~\onlinecite{Kasahara2017} using $H_{\mathrm{c}2}=\Phi_0/(2\pi \xi_0)$ with the flux quant $\Phi_0$.


	\begin{figure*}[ht]
		\centering
		\includegraphics[clip,width=1\textwidth]{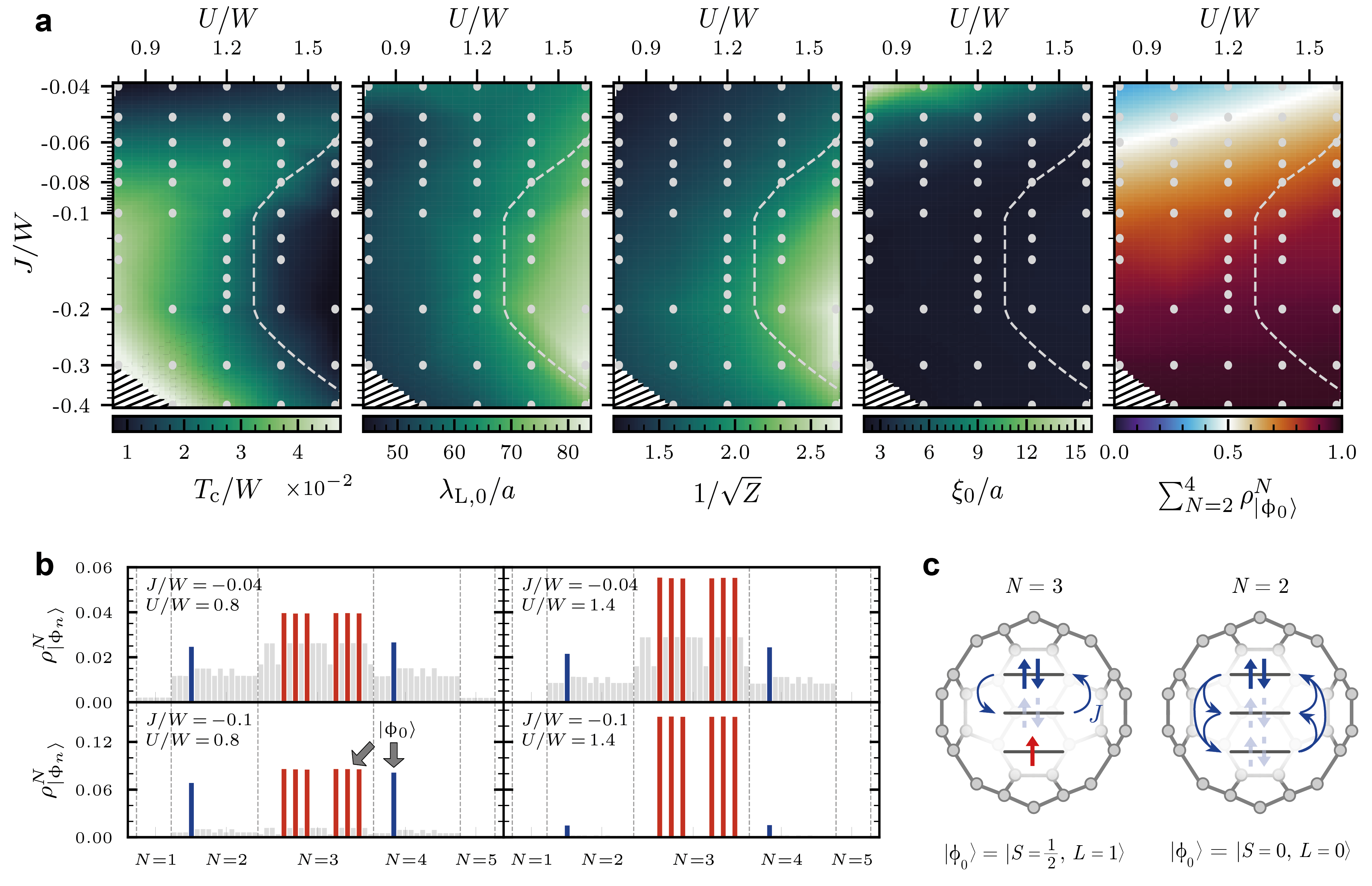}
		\caption{\textbf{Superconducting state of the \ACchem model in the ($U$, $J$)-interaction space.} (\textbf{a}) Critical temperature \Tc, zero temperature penetration depth $\lambda_{\mathrm{L},0}$, inverse square root of quasiparticle weight $Z$, coherence length $\xi_0$, and the statistical weight $\rho_{\left\vert\upphi_0\right\rangle}^N$ of the local lowest energy states $\left\vert\upphi_0\right\rangle$ of the $N = 2, 3, 4$ particle sectors obeying inverted Hund’s rules as a function of $U$ and $J$. Gray dots show original data points used for interpolation and the dashed line indicates a region where the proximity to the Mott state leads to a suppression of the superconducting state. There is no data point at the charge degeneracy line $U=2|J|$ in the lower left corner as marked by black `canceling' lines. (\textbf{b}) Distribution of statistical weights $\rho_{\left\vert\upphi_n\right\rangle}^N$ at four different interaction values $U$ and $J$. (See Supplementary Note 7 for a listing of the eigenstates $\left\vert\upphi_n\right\rangle$ and their respective local eigenenergies $E_n^N$). Red (blue) bars denote the density matrix weight of ground states in the $N = 3$ ($N = 2, 4$) particle sector, the sum of which is plotted in the last panel of \textbf{a}. (\textbf{c}) Exemplary depiction of representative lowest inverted Hund’s rule eigenstates. A delocalized doublon (electron pair) fluctuates between different orbitals due to correlated pair hopping $J$.}
		\label{Fig4} 
	\end{figure*}
	
	\subsection*{Circumvention of the lattice BCS--BEC crossover upon boosting inverted Hund’s coupling}
	The inverted Hund's coupling is crucial for superconductivity in \ACchem. This premise motivates us to explore in Fig.~\ref{Fig4} the nature of the SC state in the interaction ($U$,\,$J$) phase space for $J<0$ beyond \textit{ab initio} estimates.
	
	As long as $|J|<U/2$, we find that strengthening the negative Hund's coupling enhances the SC critical temperature with an increase up to $\Tc\approx 5\times10^{-2}$\,$W$, i.e., by a factor of seven compared to the \textit{ab initio} motivated case of $J/W=-0.04$. There is, however, a change in the role that $U$ plays in the formation of superconductivity. While $U$ was supportive for small magnitudes $|J|\lesssim0.05$\,$W$, it increasingly becomes unfavorable for $|J|>0.05$\,$W$ where \Tc is reduced with increasing $U$. The effect is largest close to the MI phase where superconductivity is strongly suppressed. We indicate this proximity region by a dashed line (c.f.~Supplementary Note 5-C).
	
\begin{figure}[t]
		\centering
		\includegraphics[clip,width=1\columnwidth]{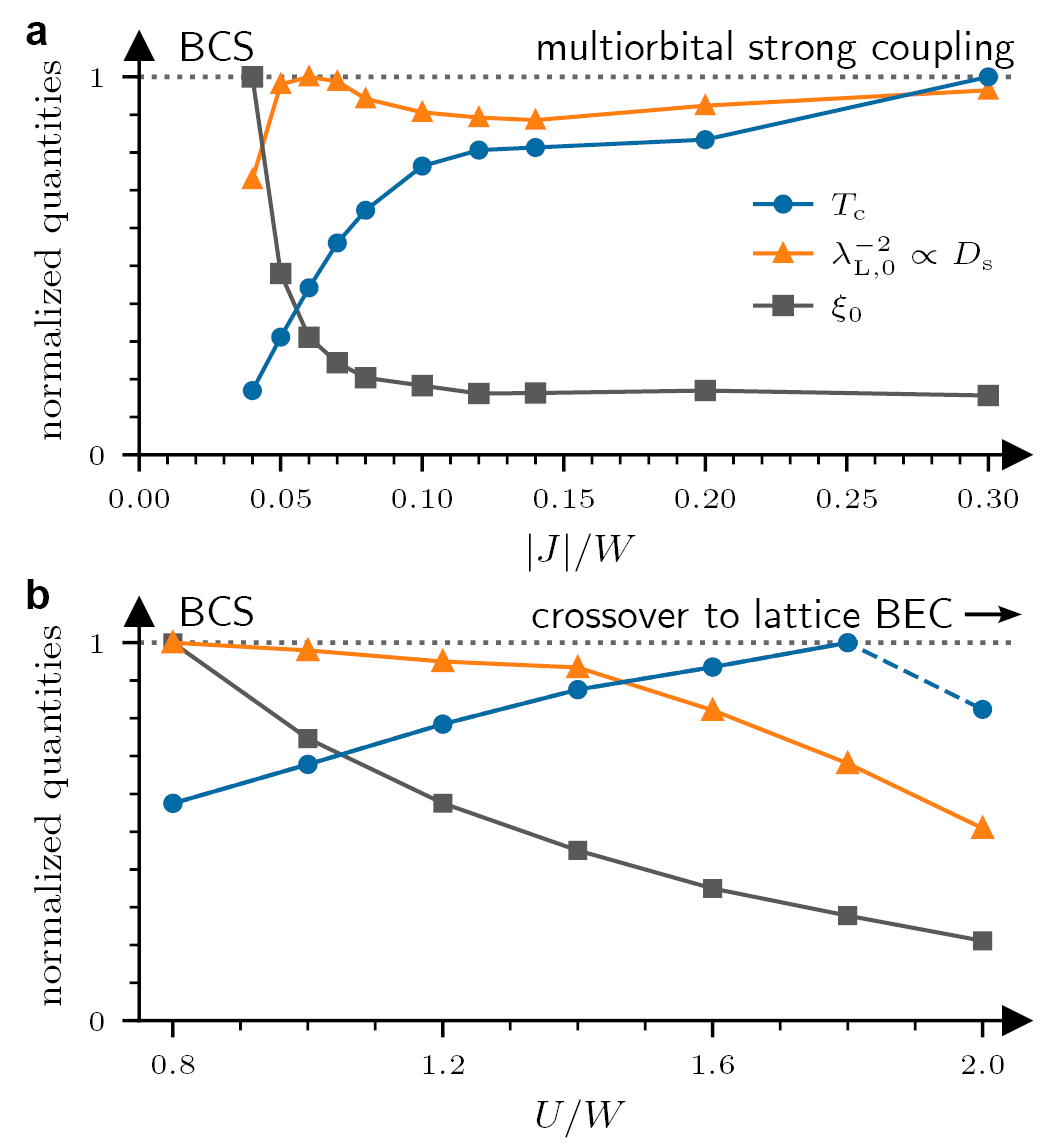}
		\caption{\textbf{Crossover of superconducting properties between different interaction regimes in the \ACchem model.} Critical temperature \Tc, coherence length $\xi_0$, and stiffness $D_{\mathrm{s}}\propto\lambda_{\mathrm{L},0}^{-2}$ when approaching the multiorbital strong coupling state as function of $|J|/W$ at fixed $U/W = 0.8$ (\textbf{a}) and when approaching the Mott insulating state as function of $U/W$ at fixed $J/W = -0.04$ (\textbf{b}). To indicate a general trend, each quantity is normalized to its maximal value within the line cut. The phenomenology of the SC regimes resembles the BCS to lattice BEC crossover when approaching the Mott insulator (\textbf{b}), which is distinct from the multiorbital strong coupling case (\textbf{a}) (c.f.~summary listed in Table~\ref{tab:SC_phases}). Note that the \Tc at $U/W = 2$ in (\textbf{b}) corresponds to the transition from a Mott insulating state instead of a metallic phase, as visually differentiated by a dashed connecting line.}
		\label{Fig5} 
	\end{figure}

	The impact of $U$ on the SC state can be understood from the $U$-dependence of the London penetration depth: \lamL grows monotonously with $U$ and reaches its maximum close to the MI phase. Hence, the condensate is softest in the region where Mott physics is important and it becomes stiffer at smaller $U$. We find that this fits to the behavior of the effective bandwidth $W_{\mathrm{eff}} = ZW\propto D_{\mathrm{s}}$ where the quasiparticle weight $Z$ is suppressed upon approaching the Mott phase. The behavior of $Z$ shown in Fig.~\ref{Fig4}a confirms the qualitative connection $\lamL\propto D_{\mathrm{s}}^{-1/2} \propto 1/\sqrt{Z}$ for $|J|>0.05$\,$W$. The $J$-dependence of \lamL is much weaker than the $U$-dependence, as can be seen in Fig.~\ref{Fig4}a and the corresponding line-cuts in Fig.~\ref{Fig5}.
	
	$\xi_0$, in contrast, depends strongly on  $J$. By just slightly increasing $|J|$ above the \textit{ab initio} estimate of $|J|$, the SC state becomes strongly localized with a short coherence length on the order of 2\,--\,3\,$a$. Remarkably, the small value of $\xi_0$ is independent of $U$ and thus the proximity to the MI phase. The localization of the condensate with  $\xi_0$ on the order of the lattice constant is reminiscent of a crossover to the BEC-type SC state. However, the dome-shaped behavior, characteristic of the lattice BCS--BEC crossover~\cite{Suzuki2022,Chen2024}, with decreasing \Tc in the strong coupling limit is notably absent here. Instead, \Tc still grows inside the plateau of minimal $\xi_0$ when increasing the effective pairing strength proportional to $|J|$ for fixed $U/W$ (c.f.~Fig.~\ref{Fig5}a). Only by diagonally traversing the ($U$, $J$) phase space, it is possible to suppress \Tc inside the short $\xi_0$ plateau with a dome structure as shown, e.g., in Ref.~\onlinecite{Hoshino2017}.

	The reason for this circumvention of the lattice BCS--BEC phenomenology can be understood from an analysis of the local density matrix weights $\rho_{|\upphi_n\rangle}$, where $|\upphi_n\rangle$ refers to the eigenstates of the local Hamiltonian of our DMFT auxiliary impurity problem. We show $\rho_{|\upphi_n\rangle}$ of four different points in the interaction phase space in Fig.~\ref{Fig4}b. In the region of short $\xi_0$, the local density matrix is dominated by only eight states (red and blue bars) given by the ``inverted Hund's rule'' ground states $|\upphi_0\rangle$ of the charge sectors with $N=2,3,4$ particles that are sketched in Fig.~\ref{Fig4}c. This can be seen in the last panel of Fig.~\ref{Fig4}a as the total weight of these eight states approaches one when entering the plateau of short $\xi_0$.
	
	By increasing $|J|$, the system is driven into a strong coupling phase where local singlets are formed as Cooper pair precursors~\cite{Han2003} while electronic hopping is \textit{not} inhibited. On the contrary, hopping between the $N = 2, 3, 4$ ground states is even facilitated via large negative $J$. It does so by affecting two different energy scales (c.f.~Supplementary Note 7): Enhancing $J<0$ reduces the atomic gap \mbox{$\Delta_{\mathrm{at}} = E_0^{N=4} + E_0^{N=2} - 2E_0^{N=3} = U - 2|J|$}~\cite{Georges2013} relevant to charge excitations and thereby supports hopping. A higher negative Hund’s exchange simultaneously increases the energy $\Delta E=2|J|$ necessary to break up the orbital singlets within a fixed charge sector. As a result, unpaired electrons in the $N=3$ state become more itinerant while the local Cooper pair binding strength increases. Since the hopping is not reduced, the SC stiffness is not compromised by larger $|J|$. This two-faced role or Janus effect of negative Hund’s exchange, that localizes Cooper pairs but delocalizes electrons, can be understood as a competition of a Mott and a charge disproportionated insulator giving way for a mixed-valence metallic state in between~\cite{Isidori2019}.
	
	Correspondingly, as the superconducting state at $-J/W>0.05$ relies on direct transitions between the local inverted Hund's ground states from filling $N=3$ to $N=2$ and $N=4$, the local Hubbard repulsion $U$ has to fulfill two requirements for superconductivity with appreciable critical temperatures. First, significant occupation of the $N=2$ and 4 states at half-filling requires that $U$ is not too large. Otherwise, the system turns Mott insulating (around $U/W\gtrsim 1.6$ for enhanced $|J|$) and the SC phase stiffness is reduced upon increasing $U$ towards the Mott limit. At the same time, a significant amount of statistical weight of the $N=3$ states demands that $U$ must not be too small either. We find that $\Delta_{\mathrm{at}}>0$ and thus $U>2|J|$ is necessary for robust SC pairing. At $\Delta_{\mathrm{at}}<0$ there is a predominance of $N=2$ and $4$ states, which couple kinetically only in second-order processes and which are susceptible to charge disproportionation~\cite{Isidori2019}. Our analysis shows that a sweet spot for a robust SC state with high phase stiffness and large \Tc exists for $|J|$ approaching $U/2$.

	
	\section*{Discussion}
	We summarize the overall change from a weak-coupling BCS state to the multiorbital strong coupling SC state characterized by \Tc, $\xi_0$, and $D_{\mathrm{s}}\propto\lambda_\mathrm{L}^{-2}$ upon enhancing $|J|$ in Fig.~\ref{Fig5}a and contrast it with the lattice BCS--BEC phenomenology in Table~\ref{tab:SC_phases}. Behavior as in the lattice BCS--BEC crossover can be found upon approaching the MI state (Fig.~\ref{Fig5}b), where strong local repulsion $U$ decreases $\xi_0$ and superconductivity compromises stiffness. As discussed above, we cannot observe a proper \Tc dome that is seen in the experimental phase diagram due to the MI state dominating over the SC state in our calculations for fixed $J/W=-0.04$~\cite{Zadik2015,Nomura2015a,Nomura2016}. However, an additional analysis of the SC gap $\Delta$ and coupling strength in Supplementary Note 6 indicates that our calculations are indeed in the vicinity of the crossover region. Diagonally traversing the ($U$,$J$) interaction plane by increasing $U$ for $J=-\gamma U$ ($\gamma\leq 0.5$) leads to a similar conclusion, although a \Tc dome can emerge more easily~\cite{Hoshino2017} as charge excitations controlled by $\Delta_{\mathrm{at}}$ are suppressed at a slower pace.
	
	Thus, two distinct localized SC states exist in the multiorbital model \ACchem\xspace -- one facilitated by local Hubbard repulsion $U$ and the other by enhanced inverted Hund’s coupling $|J|$. One might speculate under which conditions THz driving or more generally photoexcitation~\cite{Jotzu2023,Mitrano2016,Rowe2023,Budden2021} could enhance $|J|$ and steer \ACchem into this high-\Tc and short-$\xi_0$ strong coupling region, e.g., via quasiparticle trapping or displacesive meta-stability~\cite{Chattopadhyay2023}. Further experimental characterization via observables susceptible to changes in $\xi_0$ and \lamL, like critical fields, currents (see Supplementary Note 1-B), or thermoelectric and thermal transport coefficients~\cite{Ussishkin2002,Uemura2019}, can lash down the possibility of this scenario. Twisted bilayer graphene might be another platform to host the mechanism proposed in this work as an inverted Hund's pairing similar to that in \ACchem is currently under discussion~\cite{Cao2018,Dodaro2018,Wang2024}.
	
	On general grounds, the bypassing of the usual lattice BCS--BEC scenario via multiorbital physics is promising for optimization of superconducting materials to achieve higher critical currents or temperatures. Generally, limits of accessible \Tc are unknown with so far only a rigorous boundary existing for two-dimensional (2D) systems~\cite{Hazra2019,Shi2024}. An empirical upper bound to \Tc emerges from the Uemura classification~\cite{Uemura1989,Uemura1991,Uemura1991a,Uemura2019} which compares \Tc to the Fermi temperature $T_{\mathrm{F}}=E_{\mathrm{F}}/k_{\mathrm{B}}\propto D_{\mathrm{s}}$: the temperature $T_{\mathrm{BEC}}^{3\mathrm{D}}=T_{\mathrm{F}}/4.6$ of a three-dimensional (3D) non-interacting BEC. Most superconducting materials, including cuprate materials with the highest known \Tc at ambient conditions, however, are not even close to this boundary; typically, $\Tc/T_{\mathrm{BEC}} = 0.1$\,--\,$0.2$ for unconventional superconductors. Notable exceptions are monolayer FeSe\cite{Wang2012} ($\Tc/T_{\mathrm{BEC}}=0.43$), twisted graphitic systems\cite{Park2021,Cao2018} ($\Tc/T_{\mathrm{BEC}}\sim 0.37$\,--\,$0.57$), and 2D semiconductors at low carrier densities\cite{Nakagawa2021} ($\Tc/T_{\mathrm{BEC}}\sim0.36$\,--\,$0.56$)  which all reach close to the 2D boundary $T_{\mathrm{c,lim}}^{2\mathrm{D}}/T_{\mathrm{BEC}}^{3\mathrm{D}}\approx0.575$~\cite{Hazra2019}. Only ultracold Fermi gases can be tuned very close to the optimal $T_{\mathrm{BEC}}$~\cite{Chen2024a,Randeria2014}. 
	
	The empirical limitations of \Tc in most unconventional superconductors with rather high densities make sense from the standpoint and constraints of the lattice BCS--BEC crossover. In this picture, high densities seem favorable for reaching high \Tc as electrons can reach high intrinsic energy scales. Yet, lattice effects and their negative impact on the kinetic energy of superconducting carriers become also more pronounced, preventing \Tc to reach $T_{\mathrm{BEC}}$. Possible routes to evade these constraints include quantum geometric and hybridization related band structure effects~\cite{Toermae2022,Hofmann2022,Kuroki2005,Salasnich2019,Shanenko2022,Yue2022,Paramasivam2024,Nakata2017,Ryee2024} as well as the multiorbital Hund’s interaction effects triggering substantial localized stiffness uncovered here.

	\begin{table*}[th!]
		\centering
		\caption{\textbf{Characteristics of the BCS, crossover to lattice BEC, and multiorbital strong coupling limits.} Behavior of the critical temperature \Tc, coherence length $\xi_0$, superconducting stiffness $D_{\mathrm{s}}$, and the ratio of superconducting gap $\Delta$ to the Fermi energy $E_{\mathrm{F}}$ (coupling strength) in the BCS, crossover to lattice BEC~\cite{Chen2024a} and ``multiorbital strong coupling'' limit. We indicate in the last row where we can find the respective behavior in the interaction plane of the \ACchem model. We do not observe the deep BEC limit for large $U/W$ and small $J/W$, but signatures indicating the onset of the BCS--BEC crossover (c.f.~Figs.~\ref{Fig4}a and \ref{Fig5} as well as Supplementary Note 6).}
		\renewcommand{\arraystretch}{1.5}
		\newcommand\xrowht[2][0]{\addstackgap[.5\dimexpr#2\relax]{\vphantom{#1}}}
		\begin{tabularx}{\textwidth}{>{\hsize=.14\hsize\centering\arraybackslash}X >{\hsize=.28\hsize\centering\arraybackslash}X >{\hsize=.28\hsize\centering\arraybackslash}X >{\hsize=.28\hsize\centering\arraybackslash}X}
			\Xhline{.8pt}
			&\textbf{BCS limit} &\textbf{Crossover to lattice BEC limit} &\textbf{Multiorbital strong coupling} \\ \hline 
			
			\multirow{2}{*}{$\bm{T_{\mathrm{c}}}$} &
			\multirow{2}{*}{\shortstack[b]{Increase with pairing interaction, \\ 
					$\Tc\propto\Delta(T=0)$}} &
			\multirow{2}{*}{\shortstack[b]{Decrease with pairing interaction, \\
					$\Tc\propto D_{\mathrm{s}}$}} &
			\multirow{2}{*}{\shortstack[b]{Increase with pairing interaction, \\ $\Tc\propto\Delta,\,D_{\mathrm{s}}$}} \\ \\
			\multirow{2}{*}{{$\bm{\xi_0}$}}  
			&\multirow{2}{*}{{Long ($\xi_0\,\gg\,a$)}} 
			&\multirow{2}{*}{\shortstack[b]{Short ($\xi_0\,\gtrsim\,a)$ \\ increasing in (deep) BEC limit}}
			&\multirow{2}{*}{\shortstack[b]{Short ($\xi_0 \sim a)$ \\ constant with pairing interaction}} \\  \\
			\multirow{2}{*}{$\bm{D_{\mathrm{s}}\propto \lambda_{\mathrm{L}}^{-2}}$}
			&\multirow{2}{*}{Constant with pairing interaction}
			&\multirow{2}{*}{Decreasing with pairing interaction}
			&\multirow{2}{*}{\shortstack[b]{Increasing/constant with \\ pairing interaction}} \\  \\
			$\bm{\Delta}$\textbf{/}$\bm{E_{\mathrm{F}}}$
			& Small value ($\Delta/E_{\mathrm{F}} \ll 1$)
			& Large value ($\Delta/E_{\mathrm{F}} \gtrsim 1$)
			& Intermediate value ($\Delta/E_{\mathrm{F}} \sim \mathcal{O}(0.1)$) \\ 
			\multirow{2}{*}{\shortstack[b]{\textbf{Corresponding} \\ \textbf{($\bm{U}$, $\bm{J}$) region}}}
			& \multirow{2}{*}{\shortstack[b]{$\sim$ small $U/W$ \\$\sim$ small $|J|/W$}}
			& \multirow{2}{*}{\shortstack[b]{Onset at $\sim$ large $U/W$}}
			& \multirow{2}{*}{\shortstack[b]{$\sim$ small to intermediate $U/W$ \\ $\sim$ large $|J|/W$, $|J|/U<1/2$}}\\ \\
			\Xhline{.8pt}
		\end{tabularx}
		\label{tab:SC_phases}
	\end{table*}

	We emphasize that the framework to calculate the coherence length $\xi_0$ and the London penetration depth \lamL introduced in this work can be implemented in any Green function or density functional based approach to superconductivity without significant increase of the numerical complexity. Thus, our work opens the gate for ``in silico'' superconducting materials' optimization targeting not only \Tc but also $\xi_0$ and \lamL. On a more fundamental level, availability of $\xi_0$ and \lamL rather than \Tc alone can provide more constraints on possible pairing mechanisms through more rigorous theory-experiment comparisons, particularly in the domain of superconductors with strong electronic correlations.

	\section*{Methods}

	\subsection*{Dynamical mean-field theory under the constraint of finite-momentum pairing}
	We study the multiorbital interacting model $H_{\mathrm{kin}} + H_{\mathrm{int}}$ of A$_3$C$_{60}$ using Dynamical Mean-Field theory (DMFT) in Nambu space under the constraint of finite-momentum pairing (FMP). In DMFT, the lattice model is mapped onto a single Anderson impurity problem with a self-consistent electronic bath that can be solved numerically exactly. The self-energy becomes purely local $\Sigma_{ij}(i\omega_n)=\delta_{ij}\Sigma(i\omega_n)$ capturing all local correlation effects. To this end, we have to solve the following set of self-consistent equations
	\begin{align}
		\left\{
		\begin{aligned}
			\underline{G}_{\mathrm{loc}}(i\omega_n) &=  \frac{1}{N_{\bm{k}}} \sum_{\bm{k}}  \underline{G}_{\bm{k}}(i\omega_n)\\
			&= \frac{1}{N_{\bm{k}}} \sum_{\bm{k}} \left[(i\omega_n + \mu)\mathds{1} - \underline{h}(\bm{k}) - \underline{\Sigma}(i\omega_n) \right]^{-1}\,\\
			\underline{G}_{\mathrm{W}}^{-1}(i\omega_n) &= \underline{G}^{-1}_{\mathrm{loc}}(i\omega_n) + \underline{\Sigma}(i\omega_n)\,\\
			\underline{\Sigma}(i\omega_n) &= \underline{G}_{\mathrm{W}}^{-1}(i\omega_n) - \underline{G}_{\mathrm{imp}}^{-1}(\omega_n)
		\end{aligned}\right.
		\label{eq:DMFT_selfconsistency_cycle}
	\end{align}
	in DMFT~\cite{Georges1996}. Here, the local Green function $G_{\mathrm{loc}}$ is obtained from the lattice Green function $G_{\bm{k}}$ (first line) in order to construct the impurity problem by calculating the Weiss field $G_{\mathrm{W}}$ (second line). Solving the impurity problem yields the impurity Green function $G_{\mathrm{imp}}$ from which the self-energy $\Sigma$ is derived (third line). Underlined quantities denote matrices with respect to orbital indices ($\alpha$), $\mathds{1}$ is the unit matrix in orbital space, $\omega_n = (2n+1)\pi T$ denote Matsubara frequencies, $h_{\alpha\gamma}(\bm{k}) = \sum_{j} t_{\alpha\gamma}(\bm{R}_j)\mathrm{e}^{i\bm{k}\bm{R}_j}$ is the Fourier transform of the hopping matrix in Eq.~(\ref{eq:kinetic_energy}), $\mu$ indicates the chemical potential, and $N_{\bm{k}}$ is the number of $\bm{k}$-points in the momentum mesh. Convergence of the self-consistency problem is reached when the equality $\underline{G}_{\mathrm{loc}}(i\omega_n) = \underline{G}_{\mathrm{imp}}(i\omega_n)$ holds.
	
	We can study superconductivity directly in the symmetry-broken phase by extending the formalism to Nambu--Gor'kov space. The Nambu--Gor'kov function under the constraint of FMP (c.f.~Eq.~(\ref{eq:NG_GF})) on Matsubara frequencies
	\begin{widetext}
		\begin{align}
			[\underline{\mathcal{G}}_{\bm{q}}(i\omega_n,\bm{k})]^{-1}
			&=\begin{pmatrix}
				(i\omega_n + \mu)\mathds{1} - \underline{h}(\bm{k}+\frac{\bm{q}}{2}) - \underline{\Sigma}^{\mathrm{N}}(i\omega_n) 
				&-\underline{\Sigma}^{\mathrm{AN}}(i\omega_n)\\
				-\underline{\Sigma}^{\mathrm{AN}}(i\omega_n) 
				&(i\omega_n - \mu)\mathds{1} + \underline{h}(-\bm{k}+\frac{\bm{q}}{2}) + [\underline{\Sigma}^{\mathrm{N}}]^*(i\omega_n)
			\end{pmatrix} 
			\label{eq:Nambu_Gorkov_full_GF_FMP}
		\end{align}
	\end{widetext}
	then takes the place of the lattice Green function $\underline{G}_{\bm{k}}(i\omega_n)\mapsto\underline{\mathcal{G}}_{\bm{q}}(i\omega_n,\bm{k})$ in the self-consistency cycle in Eq.~(\ref{eq:DMFT_selfconsistency_cycle}). In addition to the normal component $\Sigma^{\mathrm{N}}\equiv\Sigma$, the self-energy gains an anomalous matrix element $\Sigma^{\mathrm{AN}}$ for which the gauge is chosen such that it is real-valued.
	
	We use a $35\times35\times35$ $\bm{k}$-mesh and 43200 Matsubara frequencies to set up the lattice Green function in the DMFT loop. In order to solve the local impurity problem, we use a continuous-time quantum Monte Carlo (CT-QMC) solver~\cite{Gull2011} based on the strong coupling expansion in the hybridization function (CT-HYB)~\cite{Werner2006}.
	Details on the implementation can be found in Refs.~\cite{Nomura2015a,Nomura2016}.
	Depending on calculation parameters ($T$,\,$U$,\,$J$) and proximity to the superconducting transition, we perform between $2.4\times10^{6}$ up to $19.2\times10^6$ Monte Carlo sweeps and use a Legendre expansion with 50 up to 80 basis functions. Some calculations very close to the onset of the superconducting phase transition (depending on $T$ or $\bm{q}$) needed more than 200 DMFT iterations until convergence. We average 10 or more converged DMFT iterations and calculate the mean value and standard deviation in order to estimate the uncertainty of the order parameter originating from the statistical noise of the QMC simulation.
	
	Further details on code implementation are given in Supplementary Note 4. It entails a comparison of finite-momentum pairing to the zero-momentum case\cite{Georges1996} in the Nambu--Gor'kov formalism, an explanation on the readjustment of the chemical potential $\mu$ to fix the electron filling, and a simplification of the Matsubara sum for calculating $\bm{j}_{\bm{q}}$ in Eq.~(\ref{eq:current_from_GF}).

	\begin{acknowledgments}
		We thank M.~Capone, A.~Cavalleri, M.~Eckstein, M.~Katsnelson, Y.~Iwasa, L.~Mathey, T.~Nomoto, G.~Rai, S.~Ryee, A.~Toschi, and P.~Werner for fruitful discussions.
		
		N.W., A.I.L., and T.W. acknowledge funding by the Cluster of Excellence `CUI: Advanced Imaging of Matter' of the Deutsche Forschungsgemeinschaft (DFG) (EXC 2056, Project ID 390715994) and the DFG research unit FOR 5249 (`QUAST', Project No.~449872909). Y.N. is grateful to funding via JSPS KAKENHI (No.~JP23H04869 and JP21H01041). R.A. acknowledges funding by JSPS KAKENHI (No.~JP24H00190) and RIKEN-TRIP initiative (Many-body Electron Systems). A.I.L. is supported by the European Research Council via Synergy Grant 854843 - FASTCORR. The authors gratefully acknowledge the computing time granted by the Resource Allocation Board under the project hhp00056 and provided on the supercomputer Lise at NHR@ZIB as part of the NHR infrastructure. Additional calculations were performed on the Physnet cluster at Universität Hamburg. We acknowledge financial support from the Open Access Publication Fund of Universität	Hamburg.
	\end{acknowledgments}

	\textbf{Author contributions:} N.W.~performed the calculations, prepared the figures and wrote the manuscript with input from all authors. Y.N.~and R.A.~provided the code for the DMFT calculations of the A$_3$C$_{60}$ model which A.I.L.~and N.W.~modified to include FMP. N.W.~and S.B.~analyzed the GL formulation of the FMP constraint. All authors discussed the results and physical implications. T.W., A.I.L., and N.W.~conceived the project. T.W.~supervised the project.
	
	\textbf{Competing interests:} All authors declare that they have no competing interests.

	\textbf{Data availability:} The data that support the findings of this study are available from the corresponding author upon reasonable request.

	\nocite{Kim2011,Heikkinen2014,Gull2014}
	\bibliography{references_C60}
\end{document}


\baselineskip16.8pt
	
	
	\begin{center}
		{\large \textit{Supplementary Information:}\\[0.2em]	
			\textbf{Bypassing the lattice BCS--BEC crossover in\\ strongly correlated superconductors through multiorbital physics}
		}\\[1.4em]
		
		{\normalsize Niklas Witt$^*$, Yusuke Nomura, Sergey Brener, Ryotaro Arita,\\ 
			Alexander I.~Lichtenstein, Tim O.~Wehling}\\[0.5em]
		{\small $^*$Corresponding author:  niklas.witt@physik.uni-hamburg.de}
	\end{center}
	{\centering 
	}


	\thispagestyle{empty}
	\tableofcontents
	\hypersetup{linkcolor=Aqua}
	\clearpage
	\setcounter{page}{1}

	
	

	
	\section{Phenomenological Ginzburg--Landau theory with finite-momentum pairing}
	The Ginzburg--Landau (GL) framework is a phenomenological (macroscopic) approach to the superconducting phase transition. Here, we illustrate with GL theory how introducing a FMP constraint gives access to the London penetration depth $\lambda_{\mathrm{L}}$, coherence length $\xi_0$, and also the depairing current $j_{\mathrm{dp}}$. The GL description with FMP has been discussed in other contexts like the superconducting diode effect [\href{https://doi.org/10.1073/pnas.2119548119}{55}] 
	and Fulde--Ferrel--Larkin--Ovchinnikov (FFLO) theory [\href{https://doi.org/10.1088/1361-6633/aaa4ad}{59}]. 
	Note that, as mentioned in the main text, we use the term `FMP' to refer exclusively to the order parameter with a helical phase variation and constant amplitude, whereas it is sometimes also used in the context of pair density waves [58, \href{https://doi.org/10.1146/annurev-conmatphys-031119-050711}{63}] which imprint an amplitude modulation on the superconducting gap.
	
	\subsection{Order parameter and supercurrent density}
\begin{figure}[tb]
		\centering
		\includegraphics[width=1\textwidth]{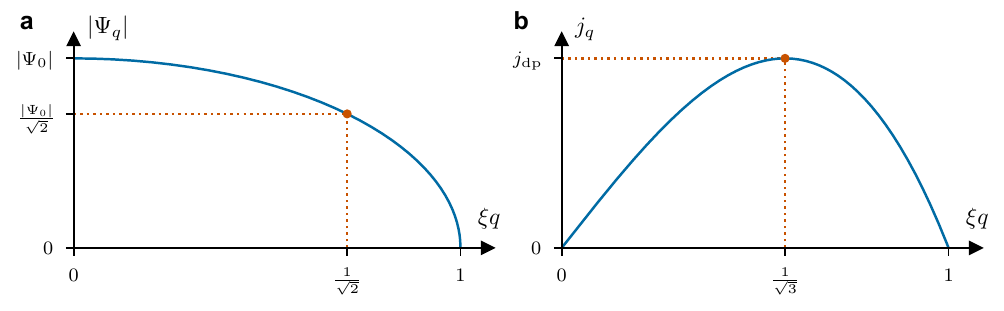}\\
		\caption{\textbf{Ginzburg--Landau solution for finite-momentum pairing.} Ginzburg--Landau expectation of the momentum ($q$)-dependent (\textbf{a}) order parameter modulus $|\Psi_q|$ and (\textbf{b}) the concomitant supercurrent density $j_q$ of a system under the constraint of finite-momentum pairing superconductivity in units of the correlation length $\xi$. We marked the characteristic length scale on which the order parameter is reduced and the point of the depairing current density $j_{\mathrm{dp}}$.}
		\label{fig:GL_OP_current_expectation}
	\end{figure}

	We start from the GL expansion of the free energy of the symmetry-broken state in terms of the complex superconducting order parameter (OP) $\Psi(\bm{r}) = |\Psi(\bm{r})|\mathrm{e}^{i\varphi(\bm{r})}$ close to the phase transition point $T_{\mathrm{c}}$ which reads
	\begin{align}
		\mathcal{F}[\Psi] = \mathcal{F}_{\mathrm{N}} + 
		\int\! \mathrm{d}^3r \left[\alpha(T) |\Psi(\boldsymbol{r})|^2 +\frac{b}{2}|\Psi(\boldsymbol{r})|^4 + \frac{\hbar^2}{2m^*} |\nabla \Psi(\boldsymbol{r})|^2\right]
		\label{eq:GL_free_energy}
	\end{align}
	where $\mathcal{F}_{\mathrm{N}}$ is the free energy of the normal state and $\alpha(T) = \alpha_0(T-T_{\mathrm{c}})$ ($\alpha_0>0$), $b>0$, and $m^*$ are the material dependent GL parameters. The GL functional encodes the two types of collective modes that emerge in the symmetry-broken state: fluctuations of the amplitude (Higgs mode) and the phase (Nambu--Goldstone mode) of the OP.
	The constraint of FMP means that we require the Cooper pairs to carry a finite fixed momentum $\bm{q}$, which translates to the requirement for the OP to be of the form $\Psi_{\bm{q}}(\bm{r}) = |\Psi_{\bm{q}}|\mathrm{e}^{i\bm{qr}}$. Then, the GL free energy density becomes
	\begin{align}
		f_{\mathrm{GL}}[\Psi_{\bm{q}}] = (\mathcal{F}[\Psi_{\bm{q}}] - \mathcal{F}_{\mathrm{N}})/V = \alpha|\Psi_{\bm{q}}|^2 + \frac{b}{2}|\Psi_{\bm{q}}|^4 + \frac{\hbar^2q^2}{2m^*} |\Psi_{\bm{q}}|^2
		\label{eq:GL_free_energy_with_q}
	\end{align}
	The gradient term in this expression has an associated length scale which is the temperature-dependent correlation length $\xi(T)$ given by
	\begin{align}
		\xi(T) = \sqrt{\frac{\hbar^2}{2m^*|\alpha|}} = \xi_0 \left(1-\frac{T}{T_{\mathrm{c}}}\right)^{-\frac{1}{2}}
		\label{eq:coherence_length}
	\end{align}
	with the coherence length $\xi_0 = \hbar/\sqrt{\alpha_0 m^*T_{\mathrm{c}}}$ at $T=0$ [4]. 
	The system's stationary point is calculated from
	\begin{align}
		\frac{\delta f_{\mathrm{GL}}}{\delta \Psi_{\bm{q}}^*} = 2\Psi_{\bm{q}} \left[\alpha(1-\xi^2q^2) + b|\Psi_{\bm{q}}|^2\right] \stackrel{\text{!}}{=} 0 \label{eq:GL_f_q}
	\end{align} 
	which results in the ${\bm{q}}$-dependence of the OP given by
	\begin{align}
		|\Psi_{\bm{q}}(T)|^2 = |\Psi_0(T)|^2 (1-\xi(T)^2q^2)
		\label{eq:GL_OP_q_dependece}
	\end{align}
	with the homogeneous OP $|\Psi_0(T)|^2 = -\alpha(T)/b \propto T-T_{\mathrm{c}}$. We plot this relation in Figure~\ref{fig:GL_OP_current_expectation}a. It shows that the OP amplitude is reduced compared to the zero-momentum pairing case for any finite $q=|\bm{q}|>0$. This suppression is induced by the nonlinear coupling of the Higgs mode to the phase mode [\href{https://doi.org/10.1146/annurev-conmatphys-031119-050813}{3}]. 
	For some critical momentum value $q_{\mathrm{c}}$, superconducting order breaks down completely ($\Psi_{q_{\mathrm{c}}}= 0$) because the kinetic energy from phase modulations exceeds the gain in energy from pairing. In GL theory, this value is given exactly by $q_{\mathrm{c}} = \xi(T)^{-1}$ (c.f.~Eq.~(\ref{eq:GL_OP_q_dependece})). The temperature dependence of the OP and extracted $\xi(T)$ gives access to the coherence length $\xi_0$ via Eq.~(\ref{eq:coherence_length}) (c.f.~Note~\ref{sec:OP_and_xi}).

	The $\bm{q}$-dependence of the OP also connects to the depairing current $j_{\mathrm{dp}}$ [\href{https://doi.org/10.1103/RevModPhys.34.667}{61}]. To see this, we derive the current density in the superconducting state. For this purpose, we (briefly) introduce a vector potential $\bm{A}$ via minimal coupling to the free energy [4, \href{https://doi.org/10.1073/pnas.2119548119}{55}]
	\begin{align}	
		\begin{split}
			f_{\mathrm{GL}}[\Psi_{\bm{q}}] &= \alpha|\Psi_{\bm{q}}|^2 + \frac{b}{2}|\Psi_{\bm{q}}|^4 + \frac{1}{2m^*}|(i\hbar\nabla + e^*\bm{A})\Psi_{\bm{q}}(\bm{r})|^2\\
			&= \alpha|\Psi_{\bm{q}}|^2 + \frac{b}{2}|\Psi_{\bm{q}}|^4 + \frac{1}{2m^*}|(\hbar^2q^2 - 4\hbar e\bm{q}\cdot\bm{A} + 4e^2A^2)|\Psi_{\bm{q}}|^2
		\end{split}
	\end{align}
	where we choose the Coulomb-gauge $\nabla\cdot\bm{A}=0$ and use $e^* = 2e$. We can explicitly set $e^*$ and, in this regard, treat the charge differently to the mass $m^*$ because $e^*$ and $(e^*)^2$ couple the same way to the OP, i.e., the charge of the Cooper pair is not renormalized. We obtain the steady-state current density from the first derivative with respect to the vector potential
	\begin{align}
		\bm{j} = \left. -\frac{\delta f_{\mathrm{GL}}}{\delta \bm{A}}\right\vert_{\bm{A}=0} =\frac{2\hbar e}{m^*}|\Psi_{\bm{q}}|^2 \left.\left(\bm{q}- \frac{2\pi}{\Phi_0}\bm{A}\right)\right\vert_{\bm{A}=0} = \frac{2\hbar e}{m^*}|\Psi_{\bm{q}}|^2 \bm{q}
		\label{eq:GL_j_derivation}
	\end{align}
	with the magnetic flux quantum $\Phi_0 = h/(2e) = \pi\hbar/e$. By inserting the OP from Eq.~(\ref{eq:GL_OP_q_dependece}), we obtain a $\bm{q}$-dependent expression for the current density ($T$-dependence suppressed)
	\begin{align}
		\bm{j}_{\bm{q}} = \frac{2\hbar e}{m^*}|\Psi_0|^2 (1-\xi^2q^2)q \, \hat{\bm{q}}
	\end{align}
	that directly shows how the Cooper pairs carry the supercurrent with their finite center-of-mass momentum along the direction $\hat{\bm{q}} = \bm{q}/q$. The current density, $j_q = |\bm{j}_q|$, is a non-monotonous function of $q$ that exhibits a maximum called depairing current, $j_{\mathrm{dp}}$ (c.f.~Figure~\ref{fig:GL_OP_current_expectation}b). $j_{\mathrm{dp}}$ can be explicitly calculated from $\partial j_q/\partial q=0$ which yields $q_{\mathrm{max}} = 1/(\sqrt{3}\xi)$ and
	\begin{align}
		j_{\mathrm{dp}} \equiv j_{q_{\mathrm{max}}}= \frac{4}{3\sqrt{3}}\frac{e\hbar |\Psi_0|^2}{m^*\xi}
		\label{eq:GL_jdp}
	\end{align}
	Since the supercurrent $\bm{j}$ is directly related to the vector potential $\bm{A}$, it is possible to derive the London equation within GL theory. We obtain the second London equation by taking the curl of Eq.~(\ref{eq:GL_j_derivation}) with $\bm{q}=0$ such that
	\begin{align}
		\frac{1}{\mu_0 \lambda_{\mathrm{L}}^2}\bm{B} = -\nabla \times \bm{j} = \frac{4\pi\hbar e|\Psi_0|^2}{m^*\Phi_0}\nabla\times\bm{A} = \frac{4e^2|\Psi_0|^2}{m^*}\bm{B}
	\end{align}
	Here, the London penetration depth $\lambda_{\mathrm{L}}$ is introduced which, in turn, can be reformulated to depend on the correlation length and depairing current:
	\begin{align}
		\lambda_{\mathrm{L}}(T) = \sqrt{\frac{m^*}{4\mu_0 e^2|\Psi_0(T)|^2}} \overset{\mathrm{Eq.}~(\ref{eq:GL_jdp})}{=} \sqrt{\frac{\Phi_0}{3\sqrt{3}\pi\mu_0\xi(T) j_{\mathrm{dp}}(T)}} = \lambda_{\mathrm{L},0}\left(1-\left(\frac{T}{T_{\mathrm{c}}}\right)^4\right)^{-\frac{1}{2}}
		\label{eq:penetration_depth}
	\end{align}
	The temperature dependence with the quartic power stated here is empirical (derived from the Gorter--Casimir model) and often used to fit experimental data [\href{https://doi.org/10.1103/PhysRev.108.1175}{39}, 40]. 
	In Note~\ref{sec:jdp_and_lamL}, we show that this temperature dependence models our DMFT data better than a linear power law as assumed in GL theory.
	
	To summarize, we obtain the correlation length $\xi(T)$ and depairing current $j_{\mathrm{dp}}(T)$ from analyzing the OP $\Psi_{\bm{q}}(\bm{r}) = |\Psi_{\bm{q}}|\mathrm{e}^{i\bm{qr}}$ subject to the FMP constraint. In a second step, we can derive the London penetration depth $\lambda_{\mathrm{L}}(T)$ from these two quantities. This connection also holds in microscopic theories, where we calculate the OP and current density from the Nambu--Gor'kov Green function (see main text and Note \ref{sec:SC_quant_calc_details}). Lastly, we want to note that the analysis of length scales $\xi(T)$ and $\lambda_{\mathrm{L}}(T)$ as done in this work is equivalent to discussing energy scales of Higgs and Nambu--Goldstone modes [\href{https://doi.org/10.1146/annurev-conmatphys-031119-050813}{3}]. 

	\subsection{Relation to experimental observables}
	$\xi$, $\lambda_{\mathrm{L}}$, and $j_{\mathrm{dp}}$ link to several experimental observables [40, \href{https://doi.org/10.1103/RevModPhys.34.667}{61}] where we suppress the $T$-dependence for brevity. 
	$j_{\mathrm{dp}}$ constitutes an upper theoretical bound to the critical current density, $j_{\mathrm{c}}$, that limits the maximal current which the superconducting state of a material can endure and which is the observable measured in experiment. The value of $j_{\mathrm{c}}$ crucially depends on sample geometry and defect densities as a current only flows near the surface shell of thickness $\sim\lambda_{\mathrm{L}}$.
	
	$\xi$ and $\lambda_{\mathrm{L}}$ are used to distinguish type I ($\xi/\lambda_{\mathrm{L}}>\sqrt{2}$) and type II ($\xi/\lambda_{\mathrm{L}}<\sqrt{2}$) superconductors and they relate to the critical magnetic fields: The first critical magnetic field
	\begin{align}
		H_{\mathrm{c}1} = \frac{\Phi_0}{4\pi\mu_0\lambda_{\mathrm{L}}^2} \ln{\frac{\lambda_{\mathrm{L}}}{\xi}}
	\end{align}
	that separates the Meissner and Abrikosov vortex lattice phases, the second critical magnetic field
	\begin{align}
		H_{\mathrm{c}2} = \frac{\Phi_0}{2\pi\mu_0\xi^2}
	\end{align} 
	that determines the magnetic field strength boundary at which the superconductor becomes a normal metal, and the thermodynamic critical field
	\begin{align}
		H_{\mathrm{c,th}} = \frac{\Phi_0}{2\sqrt{2}\pi\mu_0\xi\lambda} = \sqrt{\frac{\alpha^2}{b\mu_0}} = \sqrt{\frac{2}{\mu_0}(f_{\mathrm{N}} - f_{\mathrm{SC}})|_{\mathrm{min},q=0} }
	\end{align}

	\section{Generalized Bloch theorem in Nambu space}\label{sec:gen_Bloch_proof}
	Crystal momentum $\bm{k}$ is in general not a good quantum number for systems with spatial inhomogeneity. This applies to the situation of an arbitrary spatially varying superconducting gap $\Delta(\bm{r}) = |\Delta(\bm{r})|\mathrm{e}^{i\varphi(\bm{r})}$. In the special case of FMP with an OP of Fulde--Ferrel-type [\href{https://doi.org/10.1103/PhysRev.135.A550}{57}], 
	i.e., with helical phase $\varphi(\bm{r}) = \bm{q}\cdot\bm{r}$ and constant amplitude $|\Delta(\bm{r})|=|\Delta|$, however, a generalized lattice translation symmetry exists in Nambu space that implies a generalized Bloch theorem: We define a generalized translation operator $\mathcal{T}_n$ that acts on the Nambu spinor $\underline{\psi}^{\dagger} =  \begin{pmatrix}
		\psi^{\dagger}_{\uparrow}, &\psi^{}_{\downarrow}
	\end{pmatrix}$ of field operators $\psi^{(\dagger)}(\bm{r})$:
	\begin{align}
		\mathcal{T}_n \underline{\psi}(\bm{r}) 
		= \mathcal{T}_n \begin{pmatrix}
			\psi^{}_{\uparrow}(\bm{r})\\
			\psi^{\dagger}_{\downarrow}(\bm{r})
		\end{pmatrix} 
		= \begin{pmatrix}
			\mathrm{e}^{i\phi_n/2} \;\psi_{\uparrow}^{}(\bm{r}+\bm{R}_n)\\
			\mathrm{e}^{-i\phi_n/2} \;\;\;\psi_{\downarrow}^{\dagger}(\bm{r}+\bm{R}_n) 
		\end{pmatrix}
		= \mathrm{e}^{i\phi_n\sigma_z/2}\,\underline{\psi}(\bm{r}+\bm{R}_n)
		\label{eq:Generalized_Bloch_theorem}
	\end{align}
	Here, the spinor is not only shifted by a Bravais lattice vector $\bm{R}_n$ but it is also rotated by the angle $\phi_n = \bm{q}\cdot\bm{R}_n$ about the $z$-axis on the Bloch sphere with $\sigma_z$ being a Pauli matrix.
	
	In the following, we show that the translation defined by Eq.~(\ref{eq:Generalized_Bloch_theorem}) leaves the Hamiltonian $H=H_0 + H_{\mathrm{SC}}$ of a superconducting system consisting of lattice ($H_0$) and pairing ($H_{\mathrm{SC}}$) term invariant, i.e., that it obeys a generalized Bloch theorem. In Nambu space, the Hamiltonian takes in $d$ dimensions the form
	\begin{align}
		H_0 + H_{\mathrm{SC}} &= \int\! \mathrm{d}^d r \left[\sum_{\sigma} \left\lbrace h(\bm{r}) \psi^{\dagger}_{\sigma}(\bm{r})\psi^{}_{\sigma}(\bm{r})\right\rbrace + \Delta^{}_{\bm{q}}(\bm{r})\psi^{}_{\downarrow}(\bm{r})\psi^{}_{\uparrow}(\bm{r}) +  \Delta_{\bm{q}}^*(\bm{r})\psi^{\dagger}_{\uparrow}(\bm{r})\psi^{\dagger}_{\downarrow}(\bm{r})\right] \notag\\	
		&= \int\! \mathrm{d}^d r \left[ \underline{\psi}^{\dagger}(\bm{r}) \begin{pmatrix}
			h(\bm{r})     &\Delta_{\bm{q}}^{*}(\bm{r})\\
			\Delta^{}_{\bm{q}}(\bm{r}) & -h(\bm{r}) 
		\end{pmatrix}	\underline{\psi}(\bm{r}) \right] \notag\\
		&=\int\! \mathrm{d}^d r \;\left[\underline{\psi}^{\dagger}(\bm{r}) \left( h(\bm{r})\sigma_z + \mathrm{Re}\left\lbrace\Delta^{}_{\bm{q}}(\bm{r})\right\rbrace\sigma_x + \mathrm{Im}\left\lbrace\Delta^{}_{\bm{q}}(\bm{r})\right\rbrace\sigma_y \right)
		\underline{\psi}(\bm{r})\right]
		\label{eq:SC_Hamiltonian_Nambu_space}
	\end{align}
	with the single-particle Hamiltonian $h(\bm{r}) = -\frac{\hbar^2}{2m}\nabla^2 + V(\bm{r})$ containing the lattice periodic potential $V(\bm{r}) = V(\bm{r}+\bm{R}_n)$ and with the FMP pairing potential or gap function $\Delta^{}_{\bm{q}}(\bm{r}) = |\Delta|\mathrm{e}^{i\bm{q}\cdot\bm{r}}$. Since $\Delta\propto\Psi$, the superconducting gap carries over the phase dependence of the order parameter under the FMP constraint. From the last line of Eq.~(\ref{eq:SC_Hamiltonian_Nambu_space}), it is immediately clear that $H_0$ is invariant under translation $\mathcal{T}_n$ in Nambu space, as $\mathcal{T}_n$ trivially commutes with $h(\bm{r})\sigma_z$. The invariance of $H_{\mathrm{SC}}$ follows from the phase shift of the pairing field $\Delta_{\bm{q}}(\bm{r}+\bm{R}_n) = \Delta_{\bm{q}}(\bm{r})\mathrm{e}^{i\phi_n}$ associated with 
	\begin{align*}
		H_{\mathrm{SC}} &= \int\! \mathrm{d}^d r \left[ \underline{\psi}^{\dagger}(\bm{r}) \begin{pmatrix}
			0    &\Delta_{\bm{q}}^{*}(\bm{r})\\
			\Delta^{}_{\bm{q}}(\bm{r}) & 0
		\end{pmatrix}	\underline{\psi}(\bm{r}) \right]\\
		&=\int\! \mathrm{d}^d r \left[ \underline{\psi}^{\dagger}(\bm{r}+\bm{R}_n) \begin{pmatrix}
			0    &\Delta_{\bm{q}}^{*}(\bm{r}+\bm{R}_n)\\
			\Delta_{\bm{q}}^{}(\bm{r}+\bm{R}_n) & 0
		\end{pmatrix}	\underline{\psi}(\bm{r}+\bm{R}_n) \right]\\
		&=\int\! \mathrm{d}^d r \left[ \underline{\psi}^{\dagger}(\bm{r}+\bm{R}_n) \begin{pmatrix}
			0    &\Delta_{\bm{q}}^{*}(\bm{r})\mathrm{e}^{-i\phi_n}\\
			\Delta_{\bm{q}}^{}(\bm{r})\mathrm{e}^{i\phi_n} & 0
		\end{pmatrix}	\underline{\psi}(\bm{r}+\bm{R}_n) \right]\\
		&=\int\! \mathrm{d}^d r \left[ \underline{\psi}^{\dagger}(\bm{r}+\bm{R}_n)\, \mathrm{e}^{-i\phi_n\sigma_z/2} \begin{pmatrix}
			0    &\Delta_{\bm{q}}^{*}(\bm{r})\\
			\Delta_{\bm{q}}^{}(\bm{r})& 0
		\end{pmatrix} \mathrm{e}^{i\phi_n\sigma_z/2}\underline{\psi}(\bm{r}+\bm{R}_n) \right]\\
		&= \int\! \mathrm{d}^d r \left[ \underline{\psi}^{\dagger}(\bm{r}) \mathcal{T}_n^\dagger{} \begin{pmatrix}
			0    &\Delta_{\bm{q}}^{*}(\bm{r})\\
			\Delta_{\bm{q}}^{}(\bm{r})& 0
		\end{pmatrix} \mathcal{T}_n^{}\underline{\psi}(\bm{r}) \right]
	\end{align*}
	Thus, the generalized translation in Eq.~(\ref{eq:Generalized_Bloch_theorem}) is a symmetry of the system and (generalized) crystal momentum $\bm{k}$ constitutes a good quantum number in the case of FMP superconductivity. Note, though, that this is not true for pair density waves or generally speaking more complex FFLO-type pairings which also modulate the amplitude of the OP. In this case, methods employing supercells to accommodate for the extent of the OP modulation are necessary as was done, e.g., in Refs.~\cite{Kim2011,Heikkinen2014}.

	\section{Derivation of the supercurrent density}\label{sec:current_density_derivation_operator}
	In this section, we will derive how to calculate the charge supercurrent associated with the finite center-of-mass momentum of Cooper pairs under the FMP constraint. We start with the definition of the current operator $\bm{\hat{\jmath}}$. Generally, a current $\bm{j}$ in a system is induced by the change of the local polarization $\bm{P}$. The polarization operator is given by
	\begin{align}
		\bm{\hat{P}} = e\sum_{i} \bm{R}_i c^{\dagger}_{i} c^{}_{i} = e\sum_i \bm{R}_i n_i
	\end{align}
	for electrons of charge $e$ sitting at a lattice site $i$ (we suppress orbital and spin indices for now). The current is given by the time derivative (von-Neumann equation) of the polarization operator
	\begin{align}
		\bm{\hat{\jmath}} = \dot{\bm{\hat{P}}} = \frac{i}{\hbar}[\bm{\hat{P}},H]
		\label{eq:current_time_evolution}
	\end{align}
	We want to study the Hamiltonian with a superconducting pairing field $\Delta_{ij}$, where we here recast Eq.~(\ref{eq:SC_Hamiltonian_Nambu_space})
	\begin{align}
		H = \underbrace{\sum_{ij} t_{ij} c^{\dagger}_{i} c^{}_{j}}_{H_{\mathrm{N}}\equiv H_0} + \underbrace{\sum_{ij} \Delta_{ij}c^{}_{i}c^{}_{j}+ \Delta^*_{ij}c^{\dagger}_{j}c^{\dagger}_{i}}_{H_{\mathrm{AN}}\equiv H_{\mathrm{SC}}}
		\label{eq:full_Hamiltonian}
	\end{align}
	for discrete lattice sites $i$ instead of the continuous positions $\bm{r}$.
	To evaluate expression (\ref{eq:current_time_evolution}), we have to solve three kinds of commutators
	\begin{align*}
		[n_m, c^{\dagger}_i c_j] &= c^{\dagger}_i[n_m,c_j] + [n_m,c^{\dagger}_i]c_j = (\delta_{mi} - \delta_{mj})c^{\dagger}_i c_j\\
		[n_m, c_i c_j] &= c_i[n_m,c_j] + [n_m,c_i]c_j = -(\delta_{mi} + \delta_{mj})c_i c_j\\
		[n_m, c^{\dagger}_i c^{\dagger}_j] &= c^{\dagger}_i[n_m,c^{\dagger}_j] + [n_m,c^{\dagger}_i]c^{\dagger}_j = (\delta_{mi} + \delta_{mj})c^{\dagger}_i c^{\dagger}_j
	\end{align*}
	where we used $[A,BC] = B[A,C] + [A,B]C$ and $[n_m,c^{\dagger}_i] = \delta_{im}c^{\dagger}_i$ ($[n_m,c_i] = -\delta_{im}c_i$). We inspect the normal and anomalous component separately ($\bm{\hat{\jmath}} = \bm{\hat{\jmath}}_{\mathrm{N}} + \bm{\hat{\jmath}}_{\mathrm{AN}}$):
	\begin{align}
		\bm{\hat{\jmath}}_{\mathrm{N}} &= \frac{i}{\hbar}[\hat{\bm{P}},H_{\mathrm{N}}] = i\frac{e}{\hbar} \sum_{ijm} \bm{R}_m t_{ij} [n_m,c^{\dagger}_i c_j] = i\frac{e}{\hbar} \sum_{ijm} \bm{R}_m t_{ij}(\delta_{mi} - \delta_{mj}) c^{\dagger}_i c_j\notag\\
		&= i\frac{e}{\hbar} \sum_{ij} (\bm{R}_i - \bm{R}_j) t_{ij}c^{\dagger}_i c_j \label{eq:j_normal}\\
		\bm{\hat{\jmath}}_{\mathrm{AN}} &= \frac{i}{\hbar}[\hat{\bm{P}},H_{\mathrm{AN}}]
		= i\frac{e}{\hbar} \sum_{ijm} \bm{R}_m (\Delta_{ij} [n_m,c_i c_j] + \Delta^*_{ij} [n_m,c^{\dagger}_j c^{\dagger}_i])\notag\\
		&= -i\frac{e}{\hbar} \sum_{ijm} \bm{R}_m (\delta_{mi} + \delta_{mj})(\Delta_{ij} c_i c_j - \Delta^*_{ij} c^{\dagger}_j c^{\dagger}_i)\notag\\
		&= -i\frac{e}{\hbar} \sum_{ij} (\bm{R}_i + \bm{R}_j)(\Delta_{ij} c_i c_j -\Delta^*_{ij} c^{\dagger}_j c^{\dagger}_i)
	\end{align}
	For calculating the current density $\bm{j} = \langle{\bm{\hat{\jmath}}}\rangle$ we can make simplifications using the fact that we have local $s$-wave pairing in our system, i.e., $\Delta_{ij} \equiv \delta_{ij}\Delta\mathrm{e}^{i\bm{q}\bm{R}_i}$. Since $\langle c_i c_j\rangle  = - \langle c_j c_i\rangle$ and $\Delta_{ij} = \Delta_{ji}$, the expectation value of the anomalous part $\langle{\bm{\hat{\jmath}}_{\mathrm{AN}}}\rangle$ vanishes then.

	Since the anomalous part does not contribute, we only have to evaluate the normal component (\ref{eq:j_normal}) of the current. For this purpose, we we will assume that the states at site $i$ represent Wannier orbitals $i\to(\bm{R}_i,\alpha_i,\sigma_i)$ (orbital $\alpha$, spin $\sigma$) which are centered on the unit cell center as is the case in the A$_3$C$_{60}$ model (c.f.~Note~\ref{sec:lattice_model_current_direction}). Then, we can insert the Fourier transform of the creation and annihilation operators
	\begin{align}
		c^{}_i = \frac{1}{N_{\bm{k}}} \sum_{\bm{k}} \langle i\vert \bm{k}\rangle c^{}_{\bm{k}} = \sum_{\bm{k}} \mathrm{e}^{-i\bm{k}\bm{R}_i} \delta_{\alpha_i,\alpha_{\bm{k}}} \delta_{\sigma_i,\sigma_{\bm{k}}} c^{}_{\bm{k}}
		\;,\qquad
		c^{\dagger}_i = \frac{1}{N_{\bm{k}}} \sum_{\bm{k}} \mathrm{e}^{i\bm{k}\bm{R}_i} \delta_{\alpha_i,\alpha_{\bm{k}}} \delta_{\sigma_i,\sigma_{\bm{k}}} c^{\dagger}_{\bm{k}}
		\label{eq:ccdag_FT}
	\end{align}
	to yield
	\begin{align}
		\bm{\hat{\jmath}} = \bm{\hat{\jmath}}_{\mathrm{N}} = &i\frac{e}{\hbar} \frac{1}{N_{\bm{k}}^2}\sum_{ij\bm{k}\bm{k}^\prime} \delta_{\sigma_i,\sigma}\delta_{\sigma_j,\sigma^{\prime}}\delta_{\alpha_i,\alpha}\delta_{\alpha_j,\alpha^\prime}\delta_{\sigma_i\sigma_j}[\bm{R}_i - \bm{R}_j]t_{\alpha_i\alpha_j}(\bm{R}_i - \bm{R}_j)\mathrm{e}^{i(\bm{k}\bm{R}_i - \bm{k}^\prime \bm{R}_j)} c^{\dagger}_{\bm{k} \alpha \sigma}c^{}_{\bm{k}^\prime \alpha^\prime \sigma^\prime}\notag\\
		\overset{\bm{R}_i\mapsto\bm{R}_i + \bm{R}_j}{=} &i\frac{e}{\hbar N_{\bm{k}}}\sum_{\substack{\bm{R}_i\bm{k}\bm{k}^\prime \\ \alpha\alpha^\prime\sigma}} \bm{R}_it_{\alpha\alpha^\prime}(\bm{R}_i)\mathrm{e}^{i\bm{k}\bm{R}_i} \underbrace{\frac{1}{N_{\bm{k}}}\sum_{\bm{R}_j} \mathrm{e}^{i(\bm{k}-\bm{k}^\prime)\bm{R}_j}}_{\delta_{\bm{k}\bm{k}^\prime}} c^{\dagger}_{\bm{k} \alpha \sigma}c^{}_{\bm{k}^\prime \alpha^\prime \sigma}\notag\\
		= &\frac{e}{\hbar N_{\bm{k}}}\sum_{\bm{k}\alpha\alpha^\prime\sigma} \underbrace{i\sum_{\bm{R}_i} \bm{R}_it_{\alpha\alpha^\prime}(\bm{R}_i)\mathrm{e}^{i\bm{k}\bm{R}_i}}_{=(\nabla_{\bm{k}} h(\bm{k}))_{\alpha\alpha^\prime}} c^{\dagger}_{\bm{k} \alpha \sigma}c^{}_{\bm{k} \alpha^\prime \sigma} = \frac{e}{N_{\bm{k}}}\sum_{\bm{k}\alpha\alpha^\prime\sigma} \bm{v}_{\alpha\alpha^\prime}(\bm{k})c^{\dagger}_{\bm{k} \alpha \sigma}c^{}_{\bm{k}\alpha^\prime \sigma}
	\end{align}
	with the velocity $\bm{v}(\bm{k}) = \frac{1}{\hbar} \nabla_{\bm{k}} h(\bm{k})$. Thus, the current density is given by 
	\begin{align}
		\bm{j}_{\bm{q}} = \langle{\bm{\hat{\jmath}}}\rangle_{\bm{q}} = \frac{e}{N_{\bm{k}}}\sum_{\bm{k}\alpha\gamma\sigma} \bm{v}_{\alpha\gamma}(\bm{k})\langle c^{\dagger}_{\bm{k} \alpha \sigma}c^{}_{\bm{k} \gamma \sigma}\rangle_{\bm{q}} = \frac{2e}{N_{\bm{k}}}\sum_{\bm{k}\alpha\gamma} \bm{v}_{\alpha\gamma}(\bm{k})\langle c^{\dagger}_{\bm{k} \alpha \uparrow}c^{}_{\bm{k} \gamma \uparrow}\rangle_{\bm{q}}
		\label{eq:j_expectation_value}
	\end{align}
	where we introduced the index $\bm{q}$ to the expectation value $\langle\ldots\rangle_{\bm{q}}$ to stress that the reduced density matrix $\langle c^{\dagger}_{\bm{k} \alpha \uparrow}c^{}_{\bm{k} \gamma \uparrow}\rangle_{\bm{q}}$ is evaluated for the FMP constraint imposed on the gap and order parameter, i.e., $\Delta\mathrm{e}^{i\bm{qR_n}}$. We connect to Green function theories by writing
	\begin{align}
		\langle c^{\dagger}_{\bm{k} \alpha \uparrow}c^{}_{\bm{k} \gamma \uparrow}\rangle_{\bm{q}} = \langle c^{\dagger}_{\bm{k}-\frac{\bm{q}}{2}+\frac{\bm{q}}{2} \alpha \uparrow}c^{}_{\bm{k}-\frac{\bm{q}}{2}+\frac{\bm{q}}{2} \gamma \uparrow}\rangle_{\bm{q}} = \left[\underline{G}_{\bm{q}}\left(\tau=0^-, \bm{k}-\frac{\bm{q}}{2}\right)\right]_{\gamma\alpha} = \left[\underline{\mathcal{G}}^{\uparrow\uparrow}_{\bm{q}}\left(\tau=0^-, \bm{k}-\frac{\bm{q}}{2}\right)\right]_{\gamma\alpha}
		\label{eq:reduced_density_matrix_from_Gkio}
	\end{align}
	We want to stress that the velocity $\bm{v}(\bm{K})$ and the reduced density matrix $\langle c^{\dagger}_{\bm{K}\sigma} c^{}_{\bm{K}\sigma}\rangle$ have to carry the same momentum label $\bm{K}$ to fulfill Eq.~(\ref{eq:j_expectation_value}) (here $\bm{K}=\bm{k}$). The expression in terms of a Green function then depends decisively on the notation used for FMP in the Nambu--Gor'kov formalism. To match the definition given in Eq.~(\ref{eq:Nambu_Gorkov_GF_FMP_time}) (and Eq.~(4) of the main text) necessitates a shift of the $\bm{k}$ argument.

	Thus, we obtain an expression for the current density derived from the Nambu--Gor'kov Green function as stated in Eq.~(6) of the main text:
	\begin{align}
		\bm{j}_{\bm{q}} = \frac{2e}{N_{\bm{k}}}\sum_{\bm{k}\alpha\gamma} \bm{v}_{\alpha\gamma}(\bm{k}) \left[\underline{G}_{\bm{q}}\left(\tau=0^-, \bm{k}-\frac{\bm{q}}{2}\right)\right]_{\gamma\alpha} = \frac{2e}{N_{\bm{k}}}\sum_{\bm{k}}\mathrm{Tr}_{\alpha}\left[\underline{\bm{v}}(\bm{k}) \underline{G}_{\bm{q}}\left(\tau=0^-,\bm{k}-\frac{\bm{q}}{2}\right)\right]
		\label{eq:j_expectation_value_trace}
	\end{align}
	In practical calculations, however, we use Eq.~(\ref{eq:current_better_convergence}) as this shows better convergence with respect to the Matsubara summation associated with obtaining $G(\tau=0^{-})$. For details, see Note~\ref{sec:current_Matsubara_sum}. We note that our approach reduces to linear-response-based approaches to calculate the stiffness $D_{\mathrm{s}}\propto\lambda_{\mathrm{L}}^{-2}$ in the limit of $q\to0$. In this limit, one finds $\bm{j}=-D_{\mathrm{s}}\bm{A}$ by introducing a small $\bm{A}$ via Peierls-substitution [\href{https://doi.org/10.1103/physrevb.95.024515}{47}] which is gauge-equivalent to the FMP constraint with small $\bm{q}$ [\href{https://doi.org/10.1103/PhysRevA.8.1111}{62}]. We, however, explicitly account for finite $q$ in our calculations inside the superconducting phase and determine the depairing current $j_{\mathrm{dp}}$ for the evaluation of $\lambda_{\mathrm{L}}$ (see Note~\ref{sec:jdp_and_lamL}). We stress that one also needs the full $q$-dependence encoded in the order parameter to evaluate the correlation length $\xi(T)$.
	
	We want to note that a similar expression to Eq.~(\ref{eq:j_expectation_value_trace}) is given above Eq.~(38.13) in the book by Abrikosov, Gor'kov, and Dzyaloshinski~[60] 
	as well as in Eq.~(14.245) in the book by Coleman [4]. 
	In both cases, however, it is discussed in the context of an external magnetic field $\bm{A}$ similar to the implementation in Ref.~[\href{https://doi.org/10.1103/physrevb.95.024515}{45}].

	\section{Numerical implementation of DMFT with FMP constraint}\label{sec:SC_DMFT_calculation_details}
	
	\subsection{Nambu--Gor'kov formalism with finite-momentum pairing}
	We want to comment on the Dynamical Mean-Field theory (DMFT) calculations in Nambu space under the constraint of finite-momentum pairing (FMP). First, we summarize the description of superconductivity within DMFT using the Nambu--Gor'kov formalism entering explicitly the superconducting state for zero-momentum pairing [\href{https://doi.org/10.1103/revmodphys.68.13}{51}]. 
	Afterwards, we detail how the FMP constraint can be incorporated into the superconducting Nambu Gor'kov DMFT formalism.
	
	To extend the normal state DMFT formalism to Nambu--Gor'kov space [\href{https://doi.org/10.1103/revmodphys.68.13}{51}], 
	we perform a particle-hole transformation of the spin-down sector $c^{\dagger}_{\bm{k}\alpha\downarrow}\mapsto c^{}_{-\bm{k}\alpha\downarrow}$ and introduce Nambu spinors
	\begin{align}
		\psi^{\dagger}_{\bm{k},\alpha} = \left(c^{\dagger}_{\bm{k}\alpha\uparrow}\;\; c^{}_{-\bm{k}\alpha\downarrow}\right)\;,\quad
		\psi^{}_{\bm{k},\alpha} = \begin{pmatrix}
			c^{}_{\bm{k}\alpha\uparrow}\\ c^{\dagger}_{-\bm{k}\alpha\downarrow}
		\end{pmatrix}
		\label{eq:Nambu_spinors}
	\end{align}
	The corresponding single-particle Green function (Nambu--Gor'kov Green function) becomes a $2\times2$ matrix in Nambu space
	\begin{align}
		\mathcal{G}_{\alpha\gamma}(\tau,\bm{k}) &= -\langle T_{\tau}
		\psi^{}_{\bm{k},\alpha}(\tau)
		\psi^{\dagger}_{\bm{k},\gamma}\rangle
		= \begin{pmatrix}
			-\langle T_{\tau} c^{}_{\bm{k}\alpha\uparrow}(\tau) c^{\dagger}_{\bm{k}\gamma\uparrow}\rangle
			& -\langle T_{\tau} c^{}_{\bm{k}\alpha\uparrow}(\tau) c^{}_{-\bm{k}\gamma\downarrow}\rangle
			\\-\langle T_{\tau} c^{\dagger}_{-\bm{k}\alpha\downarrow}(\tau) c^{\dagger}_{\bm{k}\gamma\uparrow}\rangle
			& -\langle T_{\tau} c^{\dagger}_{-\bm{k}\alpha\downarrow}(\tau) c^{}_{-\bm{k}\gamma\downarrow}\rangle
		\end{pmatrix}\notag\\
		&=\begin{pmatrix}
			G_{\alpha\gamma}(\tau,\bm{k}) & F_{\alpha\gamma}(\tau,\bm{k})\\
			F^{\dagger}_{\alpha\gamma}(\tau,\bm{k}) &\bar{G}_{\alpha\gamma}(\tau,-\bm{k})\\ 
		\end{pmatrix} = 
		\begin{pmatrix}
			G_{\alpha\gamma}(\tau,\bm{k}) & F_{\alpha\gamma}(\tau,\bm{k})\\
			F^{\dagger}_{\alpha\gamma}(\tau,\bm{k}) &-G_{\alpha\gamma}(-\tau,-\bm{k})\\ 
		\end{pmatrix}
		\equiv \begin{pmatrix}
			\mathcal{G}_{\alpha\gamma}^{\uparrow\uparrow} &\mathcal{G}_{\alpha\gamma}^{\uparrow\downarrow}\\
			\mathcal{G}_{\alpha\gamma}^{\downarrow\uparrow} &\mathcal{G}_{\alpha\gamma}^{\downarrow\downarrow}
		\end{pmatrix}
		\label{eq:Nambu_Gorkov_GF_definition}
	\end{align}
	where we used $\bar{\underline{G}}(\tau,-\bm{k}) = -\underline{G}(-\tau,-\bm{k})$ for the hole propagator and $F$ denotes the anomalous (Gor'kov) Green function. The Nambu--Gor'kov Green function is determined from a Dyson equation where the non-interacting Green function
	\begin{align}
		[\underline{\mathcal{G}}^0(i\omega_n,\bm{k})]^{-1} = 
		\begin{pmatrix}
			(i\omega_n + \mu)\mathds{1} - \underline{h}(\bm{k}) &0\\
			0 & (i\omega_n - \mu)\mathds{1} + \underline{h}(-\bm{k})
		\end{pmatrix} \equiv i\omega_n\mathds{1}\cdot\sigma_0 - [\underline{h}(\bm{k}) - \mu\mathds{1}]\cdot\sigma_z
		\label{eq:Nambu_Gorkov_non-int_GF}
	\end{align}
	and self-energy
	\begin{align}
		\underline{\mathcal{S}}(i\omega_n) = \begin{pmatrix}
			\underline{\Sigma}^{\mathrm{N}}(i\omega_n) &\underline{\Sigma}^{\mathrm{AN}}(i\omega_n)\\
			\underline{\Sigma}^{\mathrm{AN}}(i\omega_n) &-[\underline{\Sigma}^{\mathrm{N}}]^*(i\omega_n)
		\end{pmatrix} \equiv \Re\underline{\Sigma}^{\mathrm{N}}(i\omega_n)\cdot\sigma_z + i\Im \underline{\Sigma}^{\mathrm{N}}(i\omega_n)\cdot\sigma_0 + \underline{\Sigma}^{\mathrm{AN}}(i\omega_n)\cdot\sigma_x
		\label{eq:Nambu_Gorkov_self-energy}
	\end{align}
	also become matrices in Nambu space which can be expressed by Pauli matrices $\sigma_i$ ($i=0,x,y,z$) for inversion symmetry $h(\bm{k}) = h(-\bm{k})$. The self-energy $\mathcal{S}$ is obtained from solving the appropriate impurity problem defined by the Weiss field $\mathcal{G}_{\mathrm{W}}$ in Nambu space and the particle-hole-transformed interaction Hamiltonion. In addition to the normal component $\Sigma^{\mathrm{N}}\equiv\Sigma$, the self-energy gains an anomalous matrix element $\Sigma^{\mathrm{AN}}$ for which the gauge is chosen such that it is real-valued, i.e., only $\sigma_x$ is involved in constructing $\mathcal{S}$. Thus, the lattice Green function in Nambu--Gor'kov space is given by
	\begin{align}
		[\underline{\mathcal{G}}(i\omega_n,\bm{k})]^{-1} &= [\underline{\mathcal{G}}^0(i\omega_n,\bm{k})]^{-1} - \underline{\mathcal{S}}(i\omega_n)\notag\\  &=\begin{pmatrix}
			(i\omega_n + \mu)\mathds{1} - \underline{h}(\bm{k}) - \underline{\Sigma}^{\mathrm{N}}(i\omega_n) 
			&-\underline{\Sigma}^{\mathrm{AN}}(i\omega_n)\\
			-\underline{\Sigma}^{\mathrm{AN}}(i\omega_n) 
			&(i\omega_n - \mu)\mathds{1} + \underline{h}(-\bm{k}) + [\underline{\Sigma}^{\mathrm{N}}]^*(i\omega_n)
		\end{pmatrix}
		\label{eq:Nambu_Gorkov_full_GF}
	\end{align}
	The self-consistency circle of DMFT (Eq.~(9) in the Methods section) generally becomes a matrix formulation in Nambu space where the lattice Green function is replaced by the Nambu--Gor'kov Green function $\underline{G}_{\bm{k}}(i\omega_n)\mapsto\underline{\mathcal{G}}(i\omega_n,\bm{k})$. We can restate the DMFT self-consistency problem in the superconducting state using calligraphic letters
	\begin{align}
		\left\{
		\begin{aligned}
			\underline{\mathcal{G}}_{\mathrm{loc}}(i\omega_n) &=  \frac{1}{N_{\bm{k}}} \sum_{\bm{k}}  \underline{\mathcal{G}}(i\omega_n,\bm{k})\,\\
			\underline{\mathcal{G}}_{\mathrm{W}}^{-1}(i\omega_n) &= \underline{\mathcal{G}}^{-1}_{\mathrm{loc}}(i\omega_n) + \underline{\mathcal{S}}(i\omega_n)\,\\
			\underline{\mathcal{S}}(i\omega_n) &= \underline{\mathcal{G}}_{\mathrm{W}}^{-1}(i\omega_n) - \underline{\mathcal{G}}_{\mathrm{imp}}^{-1}(\omega_n)
		\end{aligned}\right.
	\end{align}

	We now want to incorporate the FMP constraint into the Nambu--Gor'kov formalism. To treat the phase $\mathrm{e}^{i\bm{qr}}$ of the OP and gap in the framework of DMFT, several possibilities exist. For the simplest implementation, we introduce a phase gauge shift that cancels the momentum dependence of the order parameter $\Psi_{\bm{q}}(\bm{R}_i) = |\Psi|\mathrm{e}^{i\bm{q}\bm{R}_i} = \langle c_{i\uparrow} c_{i\downarrow} \rangle$ at site $\bm{R}_i$. By applying the transformation $c^{}_{i\alpha\sigma}\mapsto c^{}_{i\alpha\sigma}\mathrm{e}^{i\bm{q}\bm{R}_i/2}$ ($c^{\dagger}_{i,\sigma}\mapsto c^{\dagger}_{i,\sigma}\mathrm{e}^{-i\bm{q}\bm{R}_i/2}$), the hopping amplitudes $t(\bm{R}_{ij})$ are modified to yield $\tilde{t}(\bm{R}_{ij}) = t(\bm{R}_{ij})\mathrm{e}^{i\bm{q}\bm{R}_{ij}/2}$ such that the dispersion $\underline{\varepsilon}_{\bm{k}}$ obtained from diagonalizing $\underline{h}(\bm{k})$ is effectively replaced by $\underline{\varepsilon}_{\bm{k}\pm \bm{q}/2}$, i.e., the dispersion for up and down spins gets shifted by $\pm\bm{q}/2$, respectively. This shows that the introduction of FMP breaks time-reversal symmetry. The advantage of completely transferring the $\bm{q}$-dependence to the hopping matrix is that we keep the gauge freedom to choose the anomalous self-energy to be a real-valued function.
	
	To this end, we recast Eqs.~(\ref{eq:Nambu_spinors}) to (\ref{eq:Nambu_Gorkov_full_GF}) including the FMP momentum $\bm{q}$. Through the gauge transform, the Fourier transformed creation and annihilation operators pick up the $\pm\bm{q}/2$ momentum shift such that the Nambu spinors obtain an additional parametric dependence
	\begin{align}
		\psi^{\dagger}_{\bm{k},\bm{q},\alpha} = \left(c^{\dagger}_{\bm{k}+\frac{\bm{q}}{2}\alpha\uparrow}\;\; c^{}_{-\bm{k}+\frac{\bm{q}}{2}\alpha\downarrow}\right)\;,\quad
		\psi^{}_{\bm{k}+\frac{\bm{q}}{2},\alpha} = \begin{pmatrix}
			c^{}_{\bm{k}+\frac{\bm{q}}{2}\alpha\uparrow}\\ c^{\dagger}_{-\bm{k}+\frac{\bm{q}}{2}\alpha\downarrow}
		\end{pmatrix}
		\label{eq:Nambu_spinors_FMP}
	\end{align}
	The Nambu--Gor'kov Green function consequently is parameterized by $\bm{q}$ as well
	\begin{align}
		\left[\underline{\mathcal{G}}_{\bm{q}}(\tau,\bm{k})\right]_{\alpha\gamma} &= -\langle T_{\tau}
		\psi^{}_{\bm{k},\bm{q},\alpha}(\tau)
		\psi^{\dagger}_{\bm{k},\bm{q},\gamma}\rangle
		= \begin{pmatrix}
			-\langle T_{\tau} c^{}_{\bm{k}+\frac{\bm{q}}{2}\alpha\uparrow}(\tau) c^{\dagger}_{\bm{k}+\frac{\bm{q}}{2}\gamma\uparrow}\rangle
			& -\langle T_{\tau} c^{}_{\bm{k}+\frac{\bm{q}}{2}\alpha\uparrow}(\tau) c^{}_{-\bm{k}+\frac{\bm{q}}{2}\gamma\downarrow}\rangle
			\\-\langle T_{\tau} c^{\dagger}_{-\bm{k}+\frac{\bm{q}}{2}\alpha\downarrow}(\tau) c^{\dagger}_{\bm{k}+\frac{\bm{q}}{2}\gamma\uparrow}\rangle
			& -\langle T_{\tau} c^{\dagger}_{-\bm{k}+\frac{\bm{q}}{2}\alpha\downarrow}(\tau) c^{}_{-\bm{k}+\frac{\bm{q}}{2}\gamma\downarrow}\rangle
		\end{pmatrix}\notag\\
		&=\begin{pmatrix}
			\left[\underline{G}_{\bm{q}}(\tau,\bm{k})\right]_{\alpha\gamma}
			& \left[\underline{F}_{\bm{q}}(\tau,\bm{k})\right]_{\alpha\gamma}\\
			\left[\underline{F}^{\dagger}_{\bm{q}}(\tau,\bm{k})\right]_{\alpha\gamma} &\left[\underline{\bar{G}}_{\bm{q}}(\tau,-\bm{k})\right]_{\alpha\gamma}
		\end{pmatrix} 
		\equiv \begin{pmatrix}
			\left[\underline{\mathcal{G}}_{\bm{q}}^{\uparrow\uparrow}(\tau,\bm{k})\right]_{\alpha\gamma} &\left[\underline{\mathcal{G}}_{\bm{q}}^{\uparrow\downarrow}(\tau,\bm{k})\right]_{\alpha\gamma}\\
			\left[\underline{\mathcal{G}}_{\bm{q}}^{\downarrow\uparrow}(\tau,\bm{k})\right]_{\alpha\gamma} &\left[\underline{\mathcal{G}}_{\bm{q}}^{\downarrow\downarrow}(\tau,\bm{k})\right]_{\alpha\gamma}
		\end{pmatrix}
		\label{eq:Nambu_Gorkov_GF_FMP_time}
	\end{align} 
	Generally, it holds that $\underline{\mathcal{G}}_{\bm{q}}^{\uparrow\uparrow}(\tau,\bm{k}) \neq - \underline{\mathcal{G}}_{\bm{q}}^{\downarrow\downarrow}(-\tau,-\bm{k})$ for arbitrary, finite $\bm{q}$ due to the time-reversal symmetry breaking. Note that it is possible to define the Fourier transform of $c^{(\dagger)}_i$ differently such that the pairing is non-symmetric with respect to $\bm{q}$. Another often employed notation describes Cooper pairs with electrons of momenta $\bm{k}$ (in $\mathcal{G}^{\uparrow\uparrow}$) and $-\bm{k}+\bm{q}$ (in $\mathcal{G}^{\downarrow\downarrow}$) (as depicted in Fig.~2 of the main text). We here choose the symmetric notation by putting $-\frac{\bm{q}}{2}$ to both diagonal components.
	
	On Matsubara frequencies, the Nambu--Gor'kov Green function is set up via (c.f.~Eq.~(10) in the main text)
	\begin{align}
		[\underline{\mathcal{G}}_{\bm{q}}(i\omega_n,\bm{k})]^{-1}
		&=\begin{pmatrix}
			(i\omega_n + \mu)\mathds{1} - \underline{h}(\bm{k}+\frac{\bm{q}}{2}) - \underline{\Sigma}^{\mathrm{N}}(i\omega_n) 
			&-\underline{\Sigma}^{\mathrm{AN}}(i\omega_n)\\
			-\underline{\Sigma}^{\mathrm{AN}}(i\omega_n) 
			&(i\omega_n - \mu)\mathds{1} + \underline{h}(-\bm{k}+\frac{\bm{q}}{2}) + [\underline{\Sigma}^{\mathrm{N}}]^*(i\omega_n)
		\end{pmatrix} 
		\label{eq:Nambu_Gorkov_full_GF_FMP}
	\end{align}
	We note that the approach outlined here does not require the computationally more demanding use of supercells as, e.g., implemented in Refs.~\cite{Kim2011,Heikkinen2014}  
	and discussed in the review by Kinnunen et al.~[\href{https://doi.org/10.1088/1361-6633/aaa4ad}{59}] to study FFLO-type superconductivity.
	
	To induce symmetry-breaking in our calculations, we add a small pairing field $\eta = 0.1$\,meV on the off-diagonal of the Nambu--Gor'kov Green function throughout the calculation. We keep it for the whole superconducting DMFT loop because it helps stabilizing the calculations. We checked that the presence of the small, but finite $\eta$ does not change the results. In this work, we always choose $\bm{q} = q\bm{b}_1$ along the direction of a reciprocal lattice vector $\bm{b}_1$ (c.f.~Note~\ref{sec:lattice_model_current_direction}). The calculations with finite $q$ are performed in practice as in the case of $q=0$, but with $q$ becoming an additional input parameter. Calculations can be sped up by first converging the DMFT loop for $q=0$ and then using the result as a starting point for finite $q>0$ values. 
	
	\subsection{Determination of chemical potential}
	In our calculations, we adjust the chemical potential $\mu$ in each DMFT iteration in order to keep the filling of the system fixed to $\langle n\rangle_{\bm{q}} = N_{\mathrm{orb}} = 3$, i.e., the $t_{1u}$ bands of A$_3$C$_{60}$ are half-filled. The code implementation was done in Ref.~[\href{https://doi.org/10.1126/sciadv.1500568}{67}] (see also Ref.~[\href{https://doi.org/10.1088/0953-8984/28/15/153001}{68}]) for DMFT in the Nambu--Gor'kov formalism with $q=0$, but it can also be used for finite momenta. To determine the chemical potential, we solve the following equation
	\begin{align}
		\langle{n}\rangle_{\bm{q}}  &= \frac{1}{N_{\bm{k}}}\sum_{\bm{k}\alpha\sigma} \langle c^{\dagger}_{\bm{k}+\frac{\bm{q}}{2}\alpha\sigma} c^{}_{\bm{k}+\frac{\bm{q}}{2}\alpha\sigma} \rangle
		= \frac{1}{N_{\bm{k}}}\sum_{\bm{k}\alpha} \langle c^{\dagger}_{\bm{k}+\frac{\bm{q}}{2}\alpha\uparrow}c^{}_{\bm{k}+\frac{\bm{q}}{2}\alpha\uparrow}\rangle + \langle c^{\dagger}_{-\bm{k}+\frac{\bm{q}}{2}\alpha\downarrow} c^{}_{-\bm{k}+\frac{\bm{q}}{2}\alpha\downarrow} \rangle\notag\\
		&= \frac{1}{N_{\bm{k}}}\sum_{\bm{k}\alpha} \langle 1 - c^{}_{\bm{k}+\frac{\bm{q}}{2}\alpha\uparrow}c^{\dagger}_{\bm{k}+\frac{\bm{q}}{2}\alpha\uparrow}\rangle + \langle c^{\dagger}_{-\bm{k}+\frac{\bm{q}}{2}\alpha\downarrow} c^{}_{-\bm{k}+\frac{\bm{q}}{2}\alpha\downarrow}\rangle\notag\\
		&= N_{\mathrm{orb}} + \frac{1}{N_{\bm{k}}}\sum_{\bm{k}\alpha} [\underline{G}_{\bm{q}}(\tau=0^+,\bm{k}) - \bar{\underline{G}}_{\bm{q}}(\tau=0^+,-\bm{k})]_{\alpha\alpha}\notag\\
		&=  N_{\mathrm{orb}} + \frac{1}{N_{\bm{k}}}\sum_{\bm{k}\omega_n} \mathrm{Tr}_{\alpha}[\mathcal{G}^{\uparrow\uparrow}_{\bm{q}} - \mathcal{G}^{\downarrow\downarrow}_{\bm{q}}](i\omega_n,\bm{k})\mathrm{e}^{i\omega_n 0^+}
	\end{align}
	In the second step, we relabeled momentum $\bm{k} \mapsto \bm{k}+\bm{q}$ for the spin down sector. Taking the difference $\mathcal{G}^{\uparrow\uparrow}_{\bm{q}} - \mathcal{G}^{\downarrow\downarrow}_{\bm{q}}$ helps with the convergence of the Matsubara sum to evaluate the Green functions at $\tau=0^+$ since the frequency tail becomes $\mathcal{O}(1/(i\omega_n)^2)$.

	\subsection{Handling the Matsubara summation in the calculation of the current density}\label{sec:current_Matsubara_sum}
	The expression for the current density, Eq.~(\ref{eq:j_expectation_value_trace}), contains a Matsubara sum of the spin-up Nambu--Gor'kov Green function component $\mathcal{G}^{\uparrow\uparrow} = G$ to compute the reduced density matrix $\langle c^{\dagger}_{\bm{k} \alpha \uparrow}c^{}_{\bm{k} \gamma \uparrow}\rangle_{\bm{q}}$ (c.f.~Eq.~(\ref{eq:reduced_density_matrix_from_Gkio})). The Green function typically has slow convergence of Matsubara frequencies due to the $1/(i\omega_n)$-tail at large frequencies. We can achieve better convergence by including the inverse of the diagonal part of the Nambu Gor'kov Green function, i.e., the inverse of the non-interacting Green function plus the normal self-energy. Generally, we can expand the full Nambu Gor'kov Green function from Eq.~(\ref{eq:Nambu_Gorkov_full_GF_FMP}) in the isospin space in terms of Pauli matrices 
	\begin{align}
		\mathcal{\underline{\mathcal{G}}}^{-1} = \underline{g}_{0}\sigma_0 + \underline{g}_{z}\sigma_z + \underline{g}_{x}\sigma_x
	\end{align}
	We now define 
	\begin{align}
		\underline{\mathcal{G}}_{\mathrm{N}}^{-1} = \underline{g}_0 \sigma_0 + \underline{g}_z\sigma_z  \qquad\mathrm{and}\qquad \underline{\mathcal{G}}_{\Delta}^{-1} = \underline{g}_x\sigma_x
	\end{align}
	Since $\mathcal{G}_{\mathrm{N}}$ describes a time-reversal symmetric system, the following term
	\begin{align}
		\sum_{\bm{k}} \mathrm{Tr}_{\alpha} \left[ \underline{\bm{v}}(\bm{k})\underline{\mathcal{G}}^{\uparrow\uparrow}_{\mathrm{N}}\left(\tau=0^-,\bm{k} - \frac{\bm{q}}{2}\right)\right] = 0
		\label{eq:vanishing_GN}
	\end{align}
	has to vanish. Eq.~(\ref{eq:vanishing_GN}) essentially states that the charge supercurrent is induced by the superconducting condensate which only contributes to the full Nambu Green function $\mathcal{G}$ via the anomalous self-energy contained in $\mathcal{G}^{-1}_{\Delta}$ since $\underline{g}_x \equiv \underline{\Sigma}_{\mathrm{AN}}$. Thus, we can subtract Eq.~(\ref{eq:vanishing_GN}) from the current to obtain
	\begin{align}
		\bm{j} = \frac{2e}{N_{\bm{k}}}\sum_{\bm{k}} \mathrm{Tr}_{\alpha}\left[ \underline{\bm{v}}(\bm{k})\left\lbrace\underline{\mathcal{G}} - \underline{\mathcal{G}}_{\mathrm{N}}\right\rbrace^{\uparrow\uparrow}\left(\tau=0^-,\bm{k} - \frac{\bm{q}}{2}\right)\right] = \frac{2e}{N_{\bm{k}}}\sum_{\bm{k}} \mathrm{Tr}_{\alpha}\left[ \underline{\bm{v}}(\bm{k})\delta \underline{\mathcal{G}}^{\uparrow\uparrow}\left(\tau=0^-,\bm{k} - \frac{\bm{q}}{2}\right)\right]
		\label{eq:current_better_convergence}
	\end{align}
	which has a better convergence with respect to Matsubara frequencies, since $\delta \mathcal{G}^{\uparrow\uparrow} \propto 1/(i\omega_n)^3$ at large frequencies. To see the convergence behavior, we do a Taylor expansion where we focus on the isospin dependence only:
	\begin{align*}
		\delta \mathcal{G} &= \mathcal{G} - \mathcal{G}_{\mathrm{N}} = (\mathcal{G}^{-1}_{\mathrm{N}} + G^{-1}_{\Delta})^{-1} - \mathcal{G}_{\mathrm{N}}
		= \mathcal{G}_{\mathrm{N}} (\sigma_0 + \mathcal{G}^{-1}_{\Delta} \mathcal{G}_{\mathrm{N}})^{-1} - \mathcal{G}_{\mathrm{N}}\\
		&= \mathcal{G}_{\mathrm{N}}(\sigma_0 + \mathcal{G}^{-1}_{\Delta} \mathcal{G}_{\mathrm{N}} + \mathcal{G}^{-1}_{\Delta} \mathcal{G}_{\mathrm{N}} \mathcal{G}^{-1}_{\Delta} \mathcal{G}_{\mathrm{N}} + \ldots) - \mathcal{G}_{\mathrm{N}}\\
		&= \mathcal{G}_{\mathrm{N}} \mathcal{G}^{-1}_{\Delta} \mathcal{G}_{\mathrm{N}} +  \mathcal{G}_{\mathrm{N}} \mathcal{G}^{-1}_{\Delta} \mathcal{G}_{\mathrm{N}} \mathcal{G}^{-1}_{\Delta} \mathcal{G}_{\mathrm{N}} + \ldots
	\end{align*}
	The first term of the last line does not have any diagonal components since $\mathcal{G}_{\mathrm{N}} = 1/\mathcal{G}_{\mathrm{N}}^{-1} \propto \ldots\sigma_0 + \ldots \sigma_z$ and $\mathcal{G}^{-1}_{\Delta}\propto\sigma_x$ such that their product
	\begin{align}
		\mathcal{G}_{\mathrm{N}} \mathcal{G}^{-1}_{\Delta} \mathcal{G}_{\mathrm{N}}\propto \ldots\sigma_0\sigma_x + \ldots\sigma_x\sigma_z \propto \ldots\sigma_x + \ldots\sigma_y
	\end{align}
	has only off-diagonal components. Hence, the lowest order term contributing to the $\uparrow\uparrow$ component of $\delta \mathcal{G}$ is $\mathcal{O}(\mathcal{G}_{\mathrm{N}}^3 \mathcal{G}_{\Delta}^2)$ which has a $1/(i\omega_n)^3$-tail from $\mathcal{G}_{\mathrm{N}}^3$. The Taylor expansion shows that $\mathcal{G}_{\mathrm{N}}$ is the zero-order term that causes the overall $1/(i\omega_n)$-tail of $\mathcal{G}$ which we mitigate with Eq.~(\ref{eq:current_better_convergence}). An approximate expression for the current utilizing the Taylor expansion up to lowest order can be found in Eq.~(38.13) in the book by Abrikosov, Gor'kov, and Dzyaloshinski [60]. 

We compute the momentum and Matsubara summation occurring in the expression of the supercurrent density $\bm{j}_{\bm{q}}$ (c.f.~Eq.~(\ref{eq:j_expectation_value_trace}) in Note~\ref{sec:SC_DMFT_calculation_details}) using a $35^3$ $\bm{k}$-mesh and 200 Matsubara frequencies.

\subsection{Lattice model details and current direction}\label{sec:lattice_model_current_direction}
\begin{figure}[t]
	\centering
	\includegraphics[width=1\textwidth]{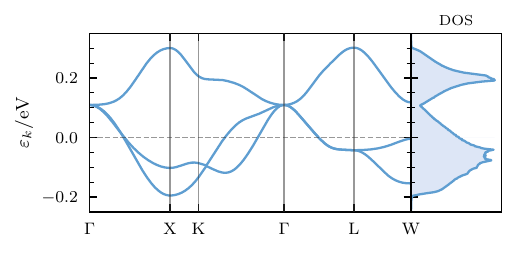}
	\caption{\textbf{Electronic structure of fullerides.} Band structure $\varepsilon_{\bm{k}}$ and density of states (DOS) of the degenerate half-filled $t_{1u}$ bands using hopping parameters for K$_3$C$_{60}$.}
	\label{fig:BS_DOS_K3C60}
\end{figure}

We here give further details on the lattice model of A$_3$C$_{60}$ materials derived in Ref.~[\href{https://doi.org/10.1103/physrevb.85.155452}{79}]. In this model, the C$_{60}$ molecules reside on a fcc lattice for which we construct Bravais lattice vectors and momenta as
\begin{align}
	\bm{R}_i = \sum_{i=1}^{3} n_j\bm{a}_j\qquad,\qquad \bm{k} = \sum_{i=j}^{3} k_j\bm{b}_j
\end{align}
with $i\equiv(n_1,n_2,n_3)$. We choose the lattice and corresponding reciprocal lattice vectors to be
\begin{align}
	& \bm{a}_1 = \frac{a}{2}(\hat{\bm{x}} + \hat{\bm{y}})\;,
	&&\bm{a}_2 = \frac{a}{2}(\hat{\bm{y}} + \hat{\bm{z}})\;, 
	&&\bm{a}_3 = \frac{a}{2}(\hat{\bm{x}} + \hat{\bm{z}})\\
	& \bm{b}_1 = \frac{2\pi}{a}( \hat{\bm{x}} + \hat{\bm{y}} - \hat{\bm{z}})\;, 
	&&\bm{b}_2 = \frac{2\pi}{a}(-\hat{\bm{x}} + \hat{\bm{y}} + \hat{\bm{z}})\;,
	&&\bm{b}_3 = \frac{2\pi}{a}( \hat{\bm{x}} - \hat{\bm{y}} + \hat{\bm{z}})
	\label{eq:lattice_vectors}
\end{align}
with lattice constant $a\sim14.2$\,--\,14.5\,\AA. $\hat{\bm{x}},\hat{\bm{y}}$, and $\hat{\bm{z}}$ are the Cartesian unit vectors. We restate the lattice model from Eq.~(8) in the main text
\begin{align}
	H_{\mathrm{kin}} = \sum_{ij}\sum_{\alpha\gamma\sigma} t_{\alpha\gamma}(\bm{R}_{ij}) c^{\dagger}_{i\alpha\sigma}c^{}_{j\gamma\sigma} = \sum_{\bm{k}}\sum_{\alpha\gamma\sigma} h_{\alpha\gamma}(\bm{k}) c^{\dagger}_{\bm{k}\alpha\sigma}c^{}_{\bm{k}\gamma\sigma}
	\label{eq:kinetic_energy}
\end{align}
where we inserted the Fourier transformation~(\ref{eq:ccdag_FT}) of creation and annihilation operators with $h_{\alpha\gamma}(\bm{k}) = \sum_{j} t_{\alpha\gamma}(\bm{R}_j)\mathrm{e}^{i\bm{k}\bm{R}_j}$. We here specify the hopping terms $t_{\alpha\gamma}(\bm{R}_{ij})$ connecting electrons of spin $\sigma$ on sites $i,j$ and molecular (Wannier) orbitals $\alpha,\gamma$ via $\bm{R}_{ij} = \bm{R}_i - \bm{R}_j$. The Wannier orbitals labeled $\alpha=1,2,3$ describe degenerate $p_x$-, $p_y$-, and $p_z$-like orbitals. For the 12 nearest-neighbor (NN) distances, the hopping matrices are given by
\begin{align*}
	&&\begin{pmatrix}
		t_1 &t_2 &0 \\
		t_2 &t_3 &0 \\
		0   &0   & t_4
	\end{pmatrix} \; \mathrm{for}\; 
	\bm{R} = (0.5,\,0.5,\,0.0), && 
	\begin{pmatrix}
		t_1  &-t_2 &0 \\
		-t_2 &t_3  &0 \\
		0    &0    & t_4
	\end{pmatrix} \; \mathrm{for}\;
	\bm{R} = (0.5,\,-0.5,\,0.0),\\
	&&\begin{pmatrix}
		t_4 &0   &0 \\
		0   &t_1 &t_2 \\
		0   &t_2 &t_3
	\end{pmatrix} \; \mathrm{for}\; 
	\bm{R} = (0,0,\,0.5,\,0.5), && 
	\begin{pmatrix}
		t_4 &0    &0 \\
		0   &t_1  &-t_2 \\
		0   &-t_2 &t_3
	\end{pmatrix} \; \mathrm{for}\; 
	\bm{R} = (0.0,\,0.5,-\,0.5),\\
	&&\begin{pmatrix}
		t_3 &0   &t_2 \\
		0   &t_4 &0 \\
		t_2 &0   & t_1
	\end{pmatrix} \; \mathrm{for}\; 
	\bm{R} = (0.5,\,0.0,\,0.5), &&  
	\begin{pmatrix}
		t_3 &0    &-t_2 \\
		0   &t_4  &0 \\
		-t_2 &0   & t_1
	\end{pmatrix} \; \mathrm{for}\; 
	\bm{R} = (-0.5,\,0.0,\,0.5)
\end{align*} 
where the connecting lattice vectors in Cartesian coordinates $\bm{R}_{ij} \equiv \bm{R} = (R_x,\,R_y,\,R_z)$ are in units of the lattice constant $a$. Hopping matrices for transfer processes to the 6 next-nearest-neighbor (NNN) sites of a C$_{60}$ molecule are
\begin{align*}
	&&\begin{pmatrix}
		t_5 &0   &0 \\
		0   &t_6 &0 \\
		0   &0   & t_7
	\end{pmatrix} \; \mathrm{for}\; 
	\bm{R} = (1,\,0,\,0), && 
	\begin{pmatrix}
		t_7 &0   &0 \\
		0   &t_5 &0 \\
		0   &0   & t_6
	\end{pmatrix} \; \mathrm{for}\;
	\bm{R} = (0,\,1,\,0), && 
	\begin{pmatrix}
		t_6 &0   &0 \\
		0   &t_7 &0 \\
		0   &0   & t_5
	\end{pmatrix} \; \mathrm{for}\;
	\bm{R} = (0,\,0,\,1)
\end{align*} 
The remaining NN and NNN hopping matrices can be generated from inversion symmetry $t_{\alpha\gamma}(\bm{R}) = t_{\alpha\gamma}(-\bm{R})$. In this work, we employ the Wannier construction for K$_3$C$_{60}$ for which the numerical values are (in meV): $t_1=-4$, $t_2=-33.9$, $t_3=42.1$, $t_4=-18.7$, $t_5=-9.3$, $t_6=-1.4$, $t_7=-0.2$. The onsite energy $t_{\alpha\alpha}(\bm{R}_{ij}=\bm{0})$ is set to zero. We show the corresponding band structure (bandwidth $W\approx0.5$\,eV) and density of states for the non-interacting model in Figure~\ref{fig:BS_DOS_K3C60}. The main difference between different A$_3$C$_{60}$ compounds is the bandwidth $W$ and effective electronic interaction strength $U$ [\href{https://doi.org/10.1088/0953-8984/28/15/153001}{68}, \href{https://doi.org/10.1103/PhysRevB.85.155452}{79}]. One can approximate the volume effect induced by different alkali dopands by changing the ratio $U/W$. The interaction Hamiltonian $H_{\mathrm{int}}$ is discussed in more detail in Note \ref{sec:Local_states}.

We also comment on the direction of the current. In all calculations, we put the FMP momentum $\bm{q}$ parallel to one of the reciprocal lattice vectors: $\bm{q}= q\bm{b}_1 = \frac{2\pi q}{a}(\hat{\bm{x}} + \hat{\bm{y}} - \hat{\bm{z}})$ such that in Cartesian coordinates $q_x = q_y = -q_z$. By this, we can employ an analytical expression for the velocity
\begin{align}
	\hbar\underline{\bm{v}}(\bm{k}) &= \nabla_{\bm{k}}\underline{h}(\bm{k}) = \nabla_{\bm{k}}\sum_{\bm{R}}\underline{t}(\bm{R})\mathrm{e}^{i\bm{k}\bm{R}}=i\sum_{\bm{R}}\bm{R}\,\underline{t}(\bm{R})\mathrm{e}^{i\bm{k}\bm{R}}
\end{align}
instead of numerically evaluating the gradient of $h(\bm{k})$ in Eq.~(\ref{eq:current_better_convergence}). The direction of the velocity and, hence, the supercurrent can be used as an internal consistency check of the code. The condition of $\bm{j}\parallel\bm{q}$ demands that
\begin{align}
	\bm{j} &= j_1 \bm{a}_1 + j_2 \bm{a}_2 + j_3 \bm{a}_3 \overset{!}{=} j(\bm{x}+\bm{y}-\bm{z})\notag\\\\
	&= a(j_1+j_3)\hat{\bm{x}} + a(j_1+j_2)\hat{\bm{y}} +  a(j_2+j_3)\hat{\bm{z}}\\
	&\Leftrightarrow j_3 = j_2 = - \frac{1}{3} j_1
\end{align}
To have the correct sign of the direction, we need $j_1 >0$ (i.e., $j_2,j_3 <0$). Since the fcc lattice has a very high symmetry, we can approximately treat the system to be isotropic. Because of this, we discuss in Fig.~2 of the main text and in Note~\ref{sec:jdp_and_lamL} only the absolute value of the current given by $|\bm{j}| = 2\sqrt{3} ax$ with $x=|j_2|=|j_3|=j_1/3$.

\section{Details on the calculation of $|\Psi_q|$,  $\xi_0$, $j_{\mathrm{dp}}$, and $\lambda_{\mathrm{L}}$ from DMFT}\label{sec:SC_quant_calc_details}
In this section, we illustrate how the order parameter $|\Psi_q|$ and concomitantly the coherence length $\xi_0$ (c.f.~Note \ref{sec:OP_and_xi}) as well as the depairing current $j_{\mathrm{dp}}$ and London penetration depth $\lambda_{\mathrm{L}}$ (c.f.~Note \ref{sec:jdp_and_lamL}) are obtained from the $\bm{q}$- and $T$-dependence of the Nambu--Gor'kov Green function in practice. Furthermore, we elaborate on how the critical temperature extracted from the temperature dependence of $\xi(T)$ and $\lambda_{\mathrm{L}}(T)$ can be used to scrutinize the proximity region of the Mott insulating phase and how it impacts the superconducting region (c.f.~Note~\ref{sec:Mott_proximity_region} and Fig.~4a of the main text).

\subsection{Order parameter and coherence length}\label{sec:OP_and_xi}
\label{sec:SC_DMFT_OP_definition}
Generally, the superconducting order parameter carries an orbital dependence. The superconducting pairing in A$_3$C$_{60}$, however, is orbital diagonal. Because of this, we perform an orbital average over the self-energy components $\Sigma^ {\mathrm{N}}$ and $\Sigma^{\mathrm{AN}}$ in each iteration step of the DMFT loop such that they are diagonal matrices in orbital space with degenerate entries ($\Sigma^{\mathrm{(A)N}}_{\alpha\gamma} = \delta_{\alpha\gamma}\,\Sigma^{\mathrm{(A)N}}$). As a result, we explicitly prevent spontaneous orbital symmetry breaking in the self-energy [\href{https://doi.org/10.1103/physrevlett.118.177002}{73}] 
and the anomalous Green function $F$ also becomes a degenerate, diagonal matrix in orbital space. This allows us to work with a single-component OP for which we take the local anomalous Green function (c.f.~Eq.~(5) of the main text)
\begin{align}
	|\Psi_{\bm{q}}| \equiv [\underline{F}^{\mathrm{loc}}_{\bm{q}}(\tau=0^-)]_{\alpha\alpha} = \sum_{\bm{k}} \langle c_{\alpha\bm{k}+\frac{\bm{q}}{2}\uparrow} c_{\alpha-\bm{k}+\frac{\bm{q}}{2}\downarrow} \rangle
\end{align}
Another option to define the OP is the superconducting gap $\Delta$, c.f.~Note \ref{sec:SC_gap}. 
Since, here, $F$ and $\Delta$ are orbital diagonal, they can be equivalently used for defining the OP as they have the same $\bm{q}$- and $T$-dependence. Taking $\Delta$ as the OP would change the relative scaling of the GL free energy because of $\Delta \approx \mathcal{U}_{\mathrm{eff}}F$ with an effective pairing potential $\mathcal{U}_{\mathrm{eff}}$ which is not of importance for determining $\xi(T)$ from the OP.

\begin{figure}[t]
	\centering
	\includegraphics[width=1\textwidth]{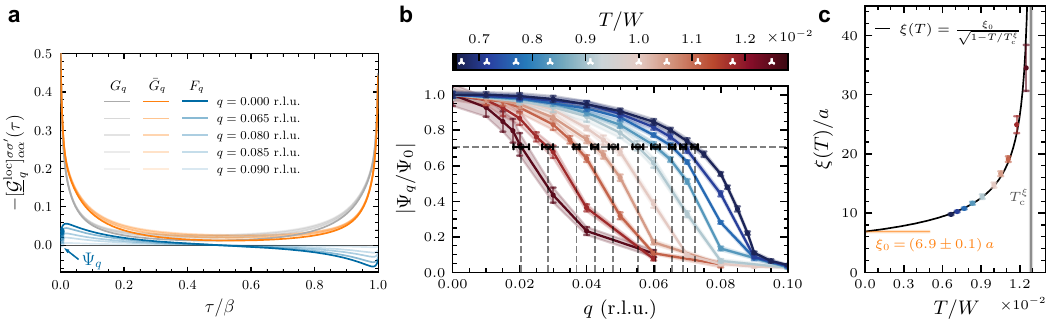}
	\caption{\textbf{Order parameter $|\Psi_q|$ and coherence length $\xi_0$ from the local anomalous Green function $F^{\mathrm{loc}}$.} (\textbf{a}) Normal (particle $G_q = \mathcal{G}_q^{\uparrow\uparrow}$, hole $\bar{G}_q = \mathcal{G}_q^{\downarrow\downarrow}$) and anomalous ($F_q = \mathcal{G}_q^{\uparrow\downarrow}$) components of the local Nambu--Gor'kov Green function $\mathcal{G}_q^{\mathrm{loc}} = \sum_{\bm{k}}\mathcal{G}_q(\bm{k})$ for different values of $\bm{q}$ at fixed $T/W=6.7\times10^{-3}$. The OP is taken at $\tau=0$ which we indicate by an arrow. (\textbf{b}) Momentum dependence of the OP normalized to the $q=0$ value for various $T$ values. The condition $|\Psi_Q(T)/\Psi_0(T)| = 1/\sqrt{2}$ to determine the correlation length via $\xi(T) = 1/(\sqrt{2}Q)$ for fixed $T$ is drawn with dashed lines. Different temperatures $T$ are indicated in the color bar by white triangular markers and the shaded areas for each $T$ show the range spanned by the uncertainty $\delta|\Psi_q|$ which we use for spline fitting to determine an error for $\xi(T)$. (\textbf{c}) Temperature dependence of $\xi(T)$ as obtained from panel \textbf{b} with the same coloring for each temperature. The fit of Eq.~(\ref{eq:coherence_length}) (Eq.~(2) in the main text) to extract $\xi_0$ is plotted with a solid black line. Shown data are results of the DMFT calculations for $U/W=1.4$ and $J/W = -0.04$, equal to the content of Figs. 2 and 3 of the main text.}
	\label{fig:OP_xi_extraction}
\end{figure}

In Figure~\ref{fig:OP_xi_extraction}a, we show the normal ($G$, $\bar{G}$) and anomalous ($F$) Green functions on imaginary time for different values of $q = |\bm{q}|$ where we also indicate the point of taking $|\Psi_q|$ at $\tau=0^+$. The amplitude of $F$ is reduced by increasing $q$, whereas $G$ and $\bar{G}$ change only slightly. Interestingly, the anomalous Green function is a \mbox{non-monotonous function of $\tau$.}

In the main text, we discuss the momentum-dependence of the OP obtained in DMFT calculations under the constraint of FMP. Here, we want to further elaborate on how $\xi(T)$ is obtained from $\Psi_{\bm{q}}(T)$. In GL theory, we found that $\xi(T) = q_{\mathrm{c}}^{-1}$ for $\lim\limits_{q\to q_{\mathrm{c}}}\Psi_q(T) = 0$. Since the point where $\Psi_{q}$ goes to zero is difficult to evaluate numerically, we use in our DMFT calculations the criterion $|\Psi_{\bm{Q}}(T)/\Psi_0(T)| = 1/\sqrt{2}$ deriving from Eq.~(\ref{eq:GL_OP_q_dependece}) such that $\xi(T)=1/(\sqrt{2}|\bm{Q}|)$ for fixed $T$. Below, we test the robustness of this criterion by comparing it to other methods of numerically extracting $\xi(T)$ from $\Psi_{\bm{q}}$. In Figure~\ref{fig:OP_xi_extraction}b, we illustrate how this criterion is applied to the DMFT results. Since our microscopic calculations include higher order terms of the free energy, the exact momentum dependence of $\Psi_q$ differs from the GL expectation (Eq.~(\ref{eq:GL_OP_q_dependece})). Note that we take $\xi(T)$ to be isotropic due to the high symmetry of the fcc lattice. In principle, it is possible to apply FMP with $\bm{q}$ in different directions in order to consider anisotropic behavior of $\xi(T)$.

Fig.~3 in the main text and Figure~\ref{fig:OP_xi_extraction} here in the Supplementary Information show error bars for $\xi(T)$ which result from propagating the statistical QMC error of the OP to $\xi(T)$. The uncertainty in $\xi(T)$ has been estimated as follows: For every dataset $|\Psi_q(T)|$ we perform a series of spline fits where we randomly vary for each $q$ the values to be fit in the range of $[|\Psi_q|-\delta|\Psi_q|,|\Psi_q|+\delta|\Psi_q|]$ spanned by the uncertainty $\delta|\Psi_q|$ of the OP. We indicate this range by color-shaded areas in Figure~\ref{fig:OP_xi_extraction}b. Based on each spline interpolation, we obtain a value for $Q$. The error in $Q$ is then estimated as the standard deviation of $Q$ values in the so-obtained ensemble.

The temperature dependence of extracted $\xi(T)$ and their uncertainty is plotted in panel c of Figure~\ref{fig:OP_xi_extraction}. As expected from GL theory, the correlation length diverges towards the critical temperature $T_{\mathrm{c}}$ and decays to a finite value $\xi_0$ for $T\to0$. By fitting Eq.~(\ref{eq:coherence_length}) to the data, we can extract the coherence length $\xi_0$ and also obtain a value for the critical temperature $T_{\mathrm{c}}$. We discuss the utility of extracting $T_{\mathrm{c}}$ this way in Note \ref{sec:Mott_proximity_region}.

\subsubsection*{Comparison of extraction methods for $\bm{\xi(T)}$}
As we conclude this section, we want to address the robustness of our criterion for extracting $\xi(T)$ from the $q$-dependence of the order parameter. To this end, we compare different methods of evaluating $\xi(T)$ given by:
\begin{enumerate}[(i)]
	\item \textbf{Suppression of order parameter}: Identify the momentum $Q$ where $|\Psi_Q/\Psi_0| = 1/\sqrt{2}$ and use $\xi=1/(\sqrt{2}Q)$.
	\item \textbf{Curvature method}: Calculate $\xi$ from the curvature of $f_q = |\Psi_q/\Psi_0|^2 \to 1 - \xi^2 q^2$ for $q\to0$ using $\xi^2 = - (1/2) \partial^2_q f_q\vert_{q=0}$.
	\item \textbf{Ginzburg--Landau fitting}: Fit the square-root Ginzburg--Landau (GL) expression $|\Psi_q/\Psi_0| = \sqrt{1 - \xi^2 q^2}$ (to data points close to $q$) to determine $\xi$.
\end{enumerate}

\begin{figure}[th!]
	\includegraphics[width=1\textwidth]{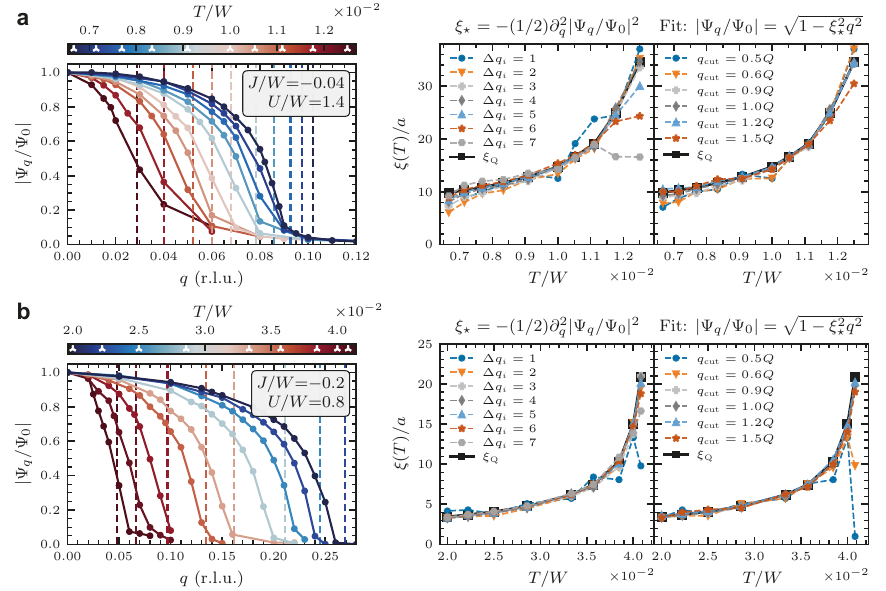}
	\caption{\textbf{Comparison of methods for extracting the coherence length $\xi$.} Panels (\textbf{a}) and (\textbf{b}) illustrate results for different values of $J/W$ and $U/W$. The left column shows the normalized order parameter $|\Psi_q/\Psi_0|$ as a function of momentum $q$ for various temperatures $T$, indicated by white triangles in the color bar. Dashed lines mark the momentum $\sqrt{2}Q = \xi_Q^{-1}$ which is associated with the coherence length $\xi$ determined by method (i), see the text for labeling of methods. The right column contrasts $\xi_Q$ (black squares) with different calculation methods for $\xi_{\star}$: (ii, left parts) evaluating the curvature of $f_q = |\Psi_q/\Psi_0|^2$ for different step sizes $\Delta q_i$, and (iii, right parts) fitting the Ginzburg--Landau expression~(\ref{eq:GL_OP_q_dependece}) for different momentum cutoffs $q_{\mathrm{cut}}$. Note that the step size $\Delta q_i$ varies across temperatures due to differing $q$-meshes, which is why only the index $i$ is listed in the legend (refer to the plots in the left column).}
	\label{fig:xi_cmp}
\end{figure}
We compare the results for $\xi(T)$ using two sets of interaction parameters in Figure~\ref{fig:xi_cmp}. The left column shows the momentum dependence of the normalized order parameter for various temperatures $T$ (similar to Figure~\ref{fig:OP_xi_extraction}b). In addition, we marked the momentum $\xi^{-1}_Q = \sqrt{2}Q$ associated with the coherence length extracted from method (i). The right column compares $\xi_Q$ (black squares) to $\xi_{\star}$ (colored points) obtained from methods (ii) and (iii) for different interval choices of numerical evaluation. For method (ii), we calculate the derivative using finite difference $\partial_q^2 f_q\vert_{q=0} \approx 2(f_{\Delta q} - f_0)/(\Delta q)^2$ and vary the step sizes $\Delta q_i = q_i - 0$ between 0.005 and 0.15 depending on $T$. Due to differing $q$-meshes for each temperature, we only list the index $i$ of the corresponding $q$-point (compare to data points in the $|\Psi_q/\Psi_0|$ plot). In approach (iii), the number of points included in the fit is controlled by $q_{\mathrm{cut}}$. We tested different cutoffs relative to the $Q$-point from method (i).

Overall, we observe strong quantitative agreement across all methods. The comparison of the two parameter sets in panels a and b reveals no systematic trend of deviations from $\xi_Q$. We find the fitting (iii) to show less variance than calculating the curvature (ii). Intricacies with numerically evaluating second derivatives from numerical data are quite common. Among the tested methods, analyzing the suppression of the order parameter at momentum $Q$ (i) has the advantage that it does not rely on fixing fitting intervals or step ranges for numerical derivatives, albeit it can pick up higher order effects in $q$. Therefore, we employ method (i) in this work.


\subsection{Current density and penetration depth}\label{sec:jdp_and_lamL}
\begin{figure}[th!]
	\centering
	\includegraphics[width=1.0\textwidth]{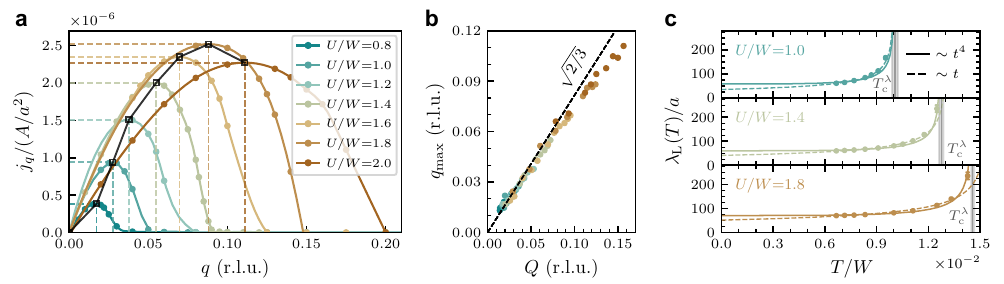}
	\caption{\textbf{Evaluation of supercurrent density $j_q = |\bm{j}_q|$ and London penetration depth $\lambda_{\mathrm{L}}$.} (\textbf{a}) Momentum-dependence of the current density for different interaction ratios $U/W$ with $J/W=-0.04$ similar to Fig.~3 of the main text and fixed $T/W=6.7\times10^{-2}$. The depairing current density $j_{\mathrm{dp}}$ (maximal $j_q$) and corresponding momentum $q_{\mathrm{max}}$ are marked by dotted lines which are extracted from a spline interpolation drawn with a solid line connecting data points. The dome-shape behavior of $j_{\mathrm{dp}}$ as a function of $U/W$ is marked by a black solid line. (\textbf{b}) Correlation between the momentum $Q$ used to calculate $\xi(T)$ from the suppression of the OP (c.f.~Note~\ref{sec:OP_and_xi}) and the momentum $q_{\mathrm{max}}$ of maximal current density $j_{\mathrm{dp}}$. Data points of the same color correspond to different temperatures for the same $U/W$ where the coloring is the same as in panel \textbf{a}. A linear function with slope $\sqrt{\frac{2}{3}}$ indicated by a dashed black line fits the data well. (\textbf{c}) Temperature dependence of the London penetration depth $\lambda_{\mathrm{L}}$ for different $U/W$ and $J/W$\,=\,$-0.04$. We plot the fit according to Eq.~(\ref{eq:penetration_depth}) (Eq.~(3) in the main text) with the quartic temperature dependence with a solid line and the fit with a linear temperature dependence with a dashed line ($t = T/T_{\mathrm{c}}^{}$).}
	\label{fig:jdp_and_lam_dependencies}
\end{figure}

We derived in Note \ref{sec:current_density_derivation_operator} an expression for the current density $\bm{j}_{\bm{q}}$ (Eq.~(\ref{eq:j_expectation_value_trace})) where we in practice employ the modified Eq.~(\ref{eq:current_better_convergence}) to ensure better convergence of the Matsubara summations. We show results of $j_q = |\bm{j}_q|$ depending on the interaction value $U/W$ for the \textit{ab initio} estimated Hund's coupling value $J/W=-0.04$ and fixed $T/W=6.7\times 10^{-2}$ in Figure~\ref{fig:jdp_and_lam_dependencies}a. $j_q$ exhibits a maximum, the depairing current $j_{\mathrm{dp}}$, that we obtain by using a spline interpolation of the calculated data. By increasing $U/W$, $j_{\mathrm{dp}}$ exhibits a dome shape which is similar to the OP but different to $T_{\mathrm{c}}$. We note that the momenta $q_{\mathrm{max}}$ where $j_{\mathrm{dp}}  = j_{q_{\mathrm{max}}}$ correlate with the momenta $Q$ used to calculate $\xi(T)$ from the OP suppression as can be seen in panel b. A line of slope $\sqrt{\frac{2}{3}}$ fits the data well suggesting $q_{\mathrm{max}} = \sqrt{\frac{2}{3}}Q$ as expected from the GL description. Only for large $U$, i.e., large values of $Q$ and $q_{\mathrm{max}}$, deviations can be seen which arise from the fact that our DMFT calculations include higher order terms which are not accounted for in the GL expansion in Eq.~(\ref{eq:GL_f_q}).

From combining the depairing current and the coherence length, we obtain the London penetration depth $\lambda_{\mathrm{L}}(T)$. In GL theory, the $T$-dependence of $\lambda_{\mathrm{L}}$ is linearized to depend on $t=T/T_{\mathrm{c}}$. However, our calculations are better described by using the empirical quartic power law $t^4$ as stated in Eq.~(\ref{eq:penetration_depth}) (Eq.~(3) of the main text). We show exemplary results of $\lambda_{\mathrm{L}}(T)$ for different $U/W$ and $J/W=-0.04$ in Figure~\ref{fig:jdp_and_lam_dependencies}c. At small $U$, the $t$ and $t^4$ dependence both match the data points quite well but the $t$-fit yields smaller values for the zero-temperature limit $\lambda_{\mathrm{L},0}$. Close to the Mott state for large $U$, the agreement becomes worse and only the $t^4$ dependence fits the data well. We observed the same behavior also in the strong coupling region for increased values of $|J|$.

\newpage
\subsection{Proximity region to the Mott transition}\label{sec:Mott_proximity_region}
From our analysis of the $T$-dependence of the zero-momentum OP $|\Psi_0|$, correlation length $\xi(T)$, and London penetration depth $\lambda_{\mathrm{L}}$, we are able to obtain different values of the superconducting transition temperature $T_{\mathrm{c}}$. In this section, we discuss how they compare and use the notation of $T_{\mathrm{c}}^{\xi,\lambda}$ to differentiate the critical temperatures obtained by fitting $\xi(T)$ and $\lambda_{\mathrm{L}}(T)$ from the $T_{\mathrm{c}}$ derived via $|\Psi_0|^2~\sim T-T_{\mathrm{c}}$.

\begin{figure}
	\centering
	\includegraphics[width=1\textwidth]{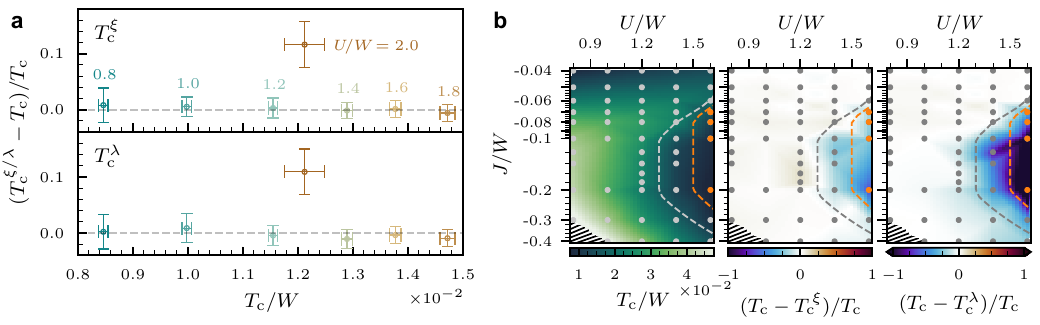}
	\caption{\textbf{Influence of proximity to the Mott insulating region on the superconducting state.} (\textbf{a}) Relative difference of critical temperature $T_{\mathrm{c}}^{(\xi,\lambda)}$ computed from fitting $\xi(T)$ and $\lambda_{\mathrm{L}}(T)$ versus $T_{\mathrm{c}}$ as obtained from the OP $|\Psi_0|$ for the data at $J/W=-0.04$ and different $U/W$ as plotted in Fig.~3 of the main text. (\textbf{b}) $T_{\mathrm{c}}$ and relative difference to $T_{\mathrm{c}}^{(\xi,\lambda)}$ as a function of interactions $U$ and $J$. Orange (gray) dots indicate data points where the critical temperature describes a transition from superconducting to Mott insulating (metallic) phase. Dashed lines are a guide to the eye separating the regions where the proximity to the Mott phase suppresses superconductivity characterized by $T_{\mathrm{c}}^{(\xi,\lambda)}<T_{\mathrm{c}}$. The $T_{\mathrm{c}}$ plot is the same as in Fig.~4a of the main text.}
	\label{fig:Tc_Mott_proximity}
\end{figure}

A first understanding can be gained by analyzing Fig.~3 of the main text. We summarize the respective critical temperatures in Figure~\ref{fig:Tc_Mott_proximity}a. Generally, the critical temperature values obtained in all three methods agree well. Only in the special case of the first-order transition from the superconducting to the Mott-insulating phase for $U/W=2$, we obtain higher values for $T_{\mathrm{c}}^{\xi,\lambda}$. We conjecture that these temperatures describe a second-order transition to a metallic state hidden by the Mott insulating phase. We can utilize this fact to gauge the influence of the Mott state to reveal a suppression of superconductivity.

In Figure~\ref{fig:Tc_Mott_proximity}b, we show the critical temperature $T_{\mathrm{c}}$ and the relative difference to $T_{\mathrm{c}}^{\xi,\lambda}$ in the ($U$,\,$J$)-plane analogous to Fig.~4a of the main text. Dots indicate original data points where orange dots (not shown in Fig.~4a) denote a critical temperature for a first-order transition from superconductor to Mott insulator. At these points, both $T_{\mathrm{c}}^\xi$ and $T_{\mathrm{c}}^\lambda$ are clearly larger than the critical temperature obtained from $|\Psi_0|^2$ which is inline with the observation at $J/W=-0.04$. However, the suppression of $T_{\mathrm{c}}$ extends to the nearby region of the direct superconductor--Mott transition. The dashed lines, of which we also draw the gray line in Fig.~4a of the main text, are a guide to the eye to separate the region where proximity to Mott insulating states leads to a suppression of the critical temperature -- even for a transition to the metallic state.

\section{Analysis of superconducting gap and coupling strength}\label{sec:SC_gap}
In this section, we analyze the superconducting gap $\Delta$ to further characterize the different superconducting regimes found in the A$_3$C$_{60}$ model. The gap is given by~\cite{Gull2014} 
\begin{align}
	\Delta(i\omega_n) = \frac{\mathrm{Re} \Sigma^{\mathrm{AN}}(i\omega_n)}{\mathds{1}-\frac{\mathrm{Im} \Sigma^{\mathrm{N}}(i\omega_n)}{\omega_n}} \equiv Z \Sigma^{\mathrm{AN}}
\end{align}
with the quasiparticle weight $Z^{-1} = \mathds{1} - \mathrm{Im}\Sigma^{\mathrm{N}}(i\omega_0)/\omega_0$ and anomalous self-energy $\Sigma^{\mathrm{AN}}$ as we evaluate the gap on the lowest Matsubara frequency $\Delta\equiv\Delta(i\omega_0)$. We note that in the (deep) BEC limit, the quasiparticle gap is not given by $\Delta$ due to the appearance of bound bosonic states. Instead, one needs to analyze $\sqrt{\Delta^2+\mu^2}$ with the chemical potential $\mu$.

In order to characterize the superconducting state, we analyze two different criteria: The first is the BCS ratio of the gap to the critical temperature $T_{\mathrm{c}}$ which in weak-coupling BCS theory has the universal value $2\Delta_0/T_{\mathrm{c}} = 3.53$ for the zero-temperature gap $\Delta_0= \Delta(T=0)$. In our calculations, we cannot reach zero temperature because of which we consider the gap for the lowest temperature $T_{\mathrm{min}}$ available for each interaction parameter set $(U, J)$. This is important for interpreting results for small Hund's coupling $J/W\lesssim-0.1$ where we could not calculate far below $T_{\mathrm{c}}$, i.e, the gap is far from saturating towards the zero-temperature value. Hence, $\Delta(T_{\mathrm{min}})$ only yields a lower bound.

\begin{figure}[t]
	\includegraphics[width=1\textwidth]{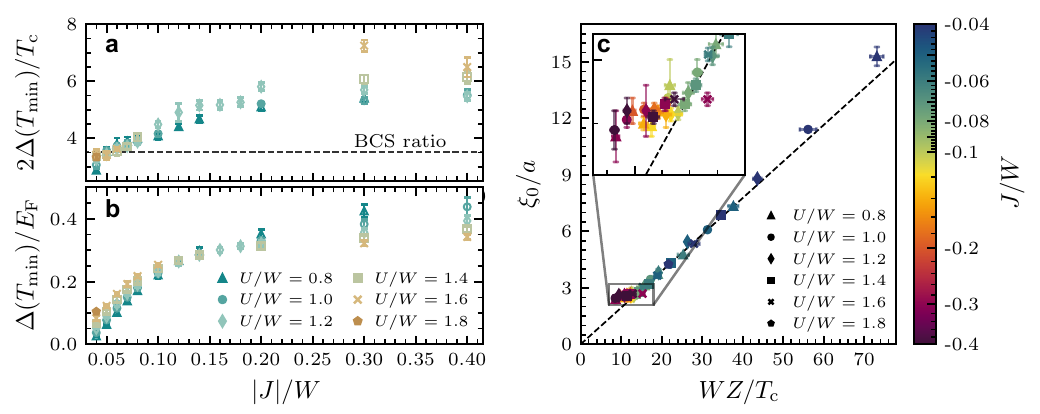}
	\caption{\textbf{Characterization of superconducting regions in the model of A$_3$C$_{60}$.} (\textbf{a}) Ratio of the superconducting gap $\Delta$ at the lowest available temperature $T_{\mathrm{min}}$ and critical temperature $T_{\mathrm{c}}$ as a function of inverted Hund's coupling $J<0$. The BCS ratio $\Delta_0/T_{\mathrm{c}} = 3.53$ is drawn with a dashed line. (\textbf{b}) Coupling strength characterized by the ratio of $\Delta$ against the Fermi energy $E_{\mathrm{F}}$. (\textbf{c}) Scaling of the coherence length $\xi_0$ with the ratio of quasiparticle weight $Z$ and $T_{\mathrm{c}}$. Note that the color scale used to mark Hund's coupling $J<0$ is logarithmic. A linear fit is drawn as a guide to the eye where the zoom-in shows  deviations for large $J<0$. We note that $T_{\mathrm{c}}$ shown in panels \textbf{a} and \textbf{c} are only for the superconductor-metal transition, not for a first-order transition to the Mott phase.}
	\label{fig:SC_gap_characterization}
\end{figure}

We show the BCS ratio in Figure~\ref{fig:SC_gap_characterization}a as a function of inverted Hund's coupling strength $J<0$ for different $U/W$. For small magnitudes $|J|/W\lesssim0.05$, our results show good agreement with the BCS value. This is in disagreement with experimental measurements [\href{https://doi.org/10.1126/sciadv.1500059}{36}, \href{https://doi.org/10.1038/ncomms14467}{80}] which observe the Cs and Rb compounds to have a ratio $2\Delta/T_{\mathrm{c}}$ much larger than the BCS value. We speculate that the discrepancy to our data arises for two reasons: First, we cannot get close to the zero-temperature value of the gap in our calculations, i.e., $T_{\mathrm{min}}$ is rather close to $T_{\mathrm{c}}$ and the ratio is likely to be underestimated by this lower bound. Second, the overestimation of $T_{\mathrm{c}}$ in DMFT can additionally lead to an underestimation of the BCS ratio. Taking into account dynamical interactions give results that are in better agreement with experiment [\href{https://doi.org/10.1126/sciadv.1500568}{67}, \href{https://doi.org/10.1088/0953-8984/28/15/153001}{68}]. Nonetheless, the qualitative trend of increasing $2\Delta/T_{\mathrm{c}}$ for larger $U/W$ fits to experimental observation. A pronounced deviation from the BCS value can be found for large inverted Hund's coupling $|J|/W>0.05$. Although $T_{\mathrm{c}}$ and $\Delta$ both increase in the ``multiorbital strong coupling'' region for enhanced $J<0$, superconductivity here is distinct to weak-coupling BCS theory. Note that we do not show $T_{\mathrm{c}}$ of a transition to the Mott insulating state.

The second criterion that we analyze is the ratio of the gap to the Fermi energy $E_{\mathrm{F}}$. This ratio $\Delta/E_{\mathrm{F}}$ can be interpreted as a dimensionless coupling strength~[\href{https://doi.org/10.1126/science.abb9860}{25}] which is small in the weak-coupling region but grows to the order of 0.1\,--\,1 in the crossover and strong-coupling regime~[\href{https://doi.org/10.1103/revmodphys.96.025002}{24}]. We note, however, that the theoretical determination of $E_{\mathrm{F}}$ is not trivial. To gauge the order of magnitude, we here resort to the non-interacting, renormalized definition 
\begin{align}
	E_{\mathrm{F}} = k_{\mathrm{B}} T_{\mathrm{F}} = \frac{\hbar^2k_{\mathrm{F}}^2}{2m^*} = \frac{\hbar^2}{2m^*} (3\pi^2n)^{\frac{2}{3}} = \frac{(3\pi^2)^{\frac{2}{3}}\hbar^2}{2m_e} \cdot Z n^{\frac{2}{3}}
\end{align}
where we inserted the quasiparticle weight $Z = m^*/m_e$. The density is $n = 3/(a/4)^3$ for the half-filled $t_{1\mathrm{u}}$ bands. We show the results in Figure~\ref{fig:SC_gap_characterization}b. For small $|J|$, the coupling strength is weak as $\Delta/E_{\mathrm{F}}< 0.1$. Towards the Mott regime, the couplings strengths grows to $~\sim0.1$, i.e., increasing $U/W$ brings the system towards the BCS--BEC crossover regime. However, increasing $|J|$ has a much stronger effect of driving the system into a strong coupling phase with $\Delta/E_{\mathrm{F}}>0.1$. Interestingly, larger $U$ here quenches the coupling strength.

Lastly, we want to characterize the superconducting regimes via the coherence length $\xi_0$. In BCS theory and Eliashberg theory, the scaling $\xi_0 \sim  v_{\mathrm{F}}^*/T_{\mathrm{c}} \propto Z/T_{\mathrm{c}}$ in terms of a renormalized Fermi velocity $v_{\mathrm{F}}^* = Zv_{\mathrm{F}}$ can be established [\href{https://doi.org/10.1103/PhysRev.108.1175}{39}--\href{https://doi.org/10.1007/bfb0120145}{41}]. 
We investigate this relation in Figure~\ref{fig:SC_gap_characterization}c. The scaling $\xi_0\sim Z/T_{\mathrm{c}}$ as in BCS and Eliashberg theory holds for most interaction values up to $J/W\sim-0.1$, even towards the Mott insulating region. It, however, deviates in the localized multiorbital strong coupling region for large $|J|/W>0.1$, where $\xi_0$ is not further reduced and $Z$ stays almost constant but $T_{\mathrm{c}}$ still increases. Thus, the multiorbital strong coupling phase is to be differentiated from a (strongly) renormalized Eliashberg system as which one might be able to describe the system in the vicinity of the Mott insulating phase.

\section{Atomic limit of three-orbital model with inverted Hund's coupling}\label{sec:Local_states}
In the main text, we found that Cooper pairs become very localized with a short coherence length $\xi_0\sim\mathcal{O}(2-3\,a)$ by increasing the inverted Hund's coupling $J<0$. It suggests that local physics become increasingly important for the formation of superconducting pairing. Indeed, this was confirmed through the analysis of local density matrix weights in the main text. Here, we want to complement the discussion of the main text with discussing the atomic limit of the interacting impurity problem.

To this end, we want to solve the Kanamori--Hubbard interaction Hamiltonian as given in Eq.~(7) of the main text without hopping processes. The form of the interaction given in the main text is convenient to read-off the (inverted) Hund's rules. We, here, restate the Kanamori--Hubbard Hamiltonian in its generalized formulation that indicates the different electronic interaction processes more clearly:
\begin{align}
	H_{\mathrm{int}} = &\sum_{\alpha} U\, n_{\alpha\uparrow} n_{\alpha\downarrow}
	+ \sum_{\alpha<\gamma,\sigma\sigma^{\prime}} (U^{\prime} - \delta_{\sigma\sigma^{\prime}}J)\, n_{\alpha\sigma}n_{\gamma\sigma^{\prime}}\notag\\
	&- \sum_{\alpha\neq\gamma}J_{\mathrm{X}}\, c^{\dagger}_{\alpha\uparrow}c^{}_{\alpha\downarrow}c^{\dagger}_{\gamma\downarrow}c^{}_{\gamma\uparrow}
	+ \sum_{\alpha\neq\gamma}J_{\mathrm{P}}\, c^{\dagger}_{\alpha\uparrow}c^{\dagger}_{\alpha\downarrow}c^{}_{\gamma\downarrow}c^{}_{\gamma\uparrow}
	\label{eq:generalized_HK}
\end{align}
It consists of intraorbital interaction $U$, interorbital interaction $U^{\prime}$, Hund's coupling $J$, spin-exchange $J_{\mathrm{X}}$, and correlated pair hopping $J_{\mathrm{P}}$. Yet, not all coupling constants are independent. We assume SU(2)$\times$SO(3) symmetry implying $J_{\mathrm{X}} = J$ and $J_{\mathrm{P}} = U - U^{\prime} - J$ [\href{https://doi.org/10.1146/annurev-conmatphys-020911-125045}{69}]. 
In the physical system and our calculations, we have in addition $J_{\mathrm{P}} = J$ resulting in $U^{\prime} = U - 2J$. In the following discussion, we will emphasize the contribution of $J_{\mathrm{P}}$ since the low-energy excitations for inverted Hund's coupling $J<0$ are only governed by  $J_{\mathrm{P}}$. It is instructive to rewrite Eq.~(\ref{eq:generalized_HK}) in the same way as Eq.~(7) of the main text (c.f.~Eq.~5 in Ref.~[\href{https://doi.org/10.1146/annurev-conmatphys-020911-125045}{69}]) to see the role of $J_{\mathrm{P}}$:
\begin{align}
	H_{\mathrm{int}} = \frac{1}{4} (2U-3J-3J_{\mathrm{P}})\hat{N}(\hat{N}-1)- (J+J_{\mathrm{P}})\hat{\bm{S}}^2 - \frac{1}{2}J_{\mathrm{P}}\hat{\bm{L}}^2 + \frac{1}{4}(3J+7J_{\mathrm{P}})\hat{N}
	\label{eq:KH_normalform}
\end{align}
The pair hopping term, most notably, dictates the energy gain from high orbital angular momentum $L^2$ and partially that of the total spin $S^2$ of a given eigenstate for this Hamiltonian. We detail the spectrum in Tab.~\ref{tab:Local_eigenstates} for the case of negative $J,J_{\mathrm{P}}<0$ and half-filled orbitals where we add a chemical potential $\mu$ to ensure particle-hole symmetry. The chemical potential at half-filling is given by $\mu = \frac{1}{2}U + \frac{N-1}{2}(2U^{\prime}-J)= \frac{5}{2}U - 3J -2J_{\mathrm{P}}$ with $N=3$ and $U^{\prime} = U-J-J_{\mathrm{P}}$, as can be inferred from particle-hole transforming Eq.~(\ref{eq:KH_normalform}). The dimension of the complete Fock space is $\mathrm{dim}\:\mathcal{H}_{\mathrm{Fock}} = 2^6 = 64$. Fig.~4b in the main text shows the statistical occupation of these 64 states during the QMC calculation in DMFT.

\begin{figure}[t]
	\centering
	\includegraphics[width=1\textwidth]{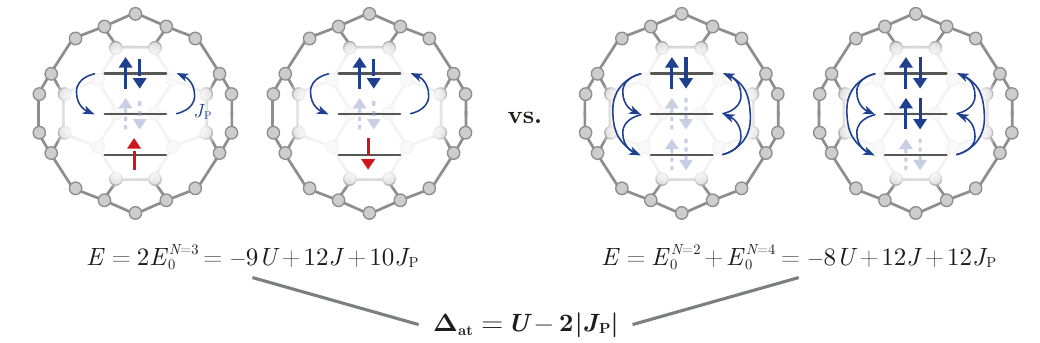}
	\caption{\textbf{Schematic energy diagram of the atomic gap $\Delta_{\mathrm{at}}$.} The left and right sides show the charge configurations (3,3) and (2,4) of neighboring C$_{60}$ molecules, respectively, whose energy difference is given by $\Delta_{\mathrm{at}}$.}
	\label{fig:atomic_gap}
\end{figure}

\begin{figure}[t]
	\centering
	\includegraphics[width=1\textwidth]{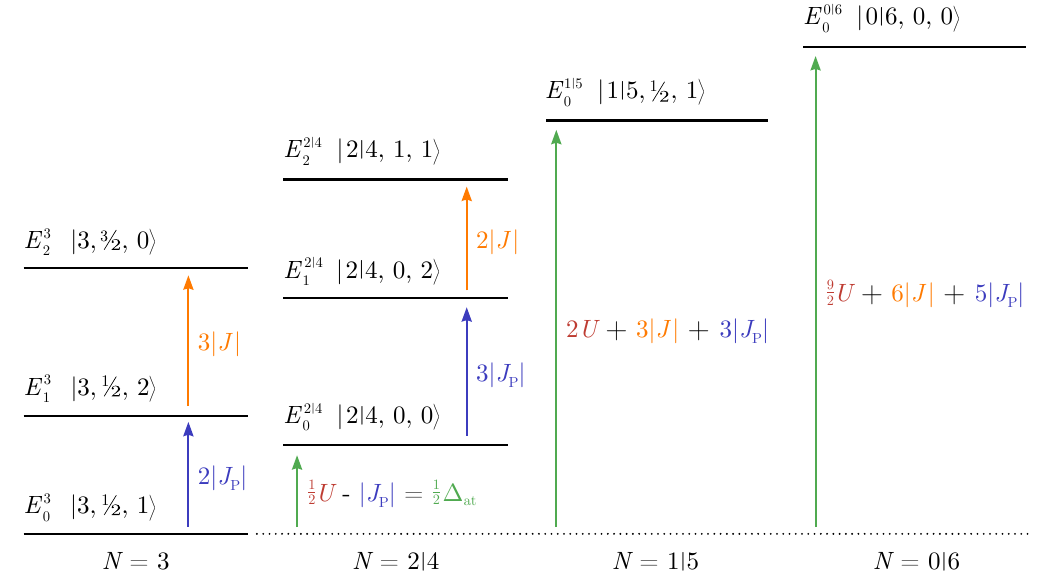}
	\caption{\textbf{Spectrum of the three-orbital Kanamori--Hubbard model.} The energies $E^N_n$ and states $\vert \upphi^N_n\rangle \equiv \vert N, S, L\rangle$ of the three-orbital atom with Kanamori--Hubbard interaction (Eq.~(\ref{eq:KH_normalform}) at half-filling correspond to the notation used in Tab.~\ref{tab:Local_eigenstates}. The relative positions of energies are drawn for the case of $U>2|J_{\mathrm{P}}|$, such that the ground state lies in the $N=3$ sector given by $E^{N=3}_0$ and $\vert N=3,S=\frac{1}{2},L=1\rangle$.}
	\label{fig:KH_spectrum}
\end{figure}

In case of half-filling, the atomic gap of the system is $\Delta_{\mathrm{at}} = E_0^{N=4} - E_0^{N=3} - (E_0^{N=3} - E_0^{N=2}) = U + 2J_{\mathrm{P}} = U - 2|J_{\mathrm{P}}|$, see Figure~\ref{fig:atomic_gap}. For $\Delta_{\mathrm{at}}>0$, i.e., $U>2|J_{\mathrm{P}}|$, the lowest energy state lies in the $N=3$ particle sector and is given by $E_0^{N=3}$.  We sketch the energy spectrum for this case in Figure~\ref{fig:KH_spectrum}. The lowest energy excitations from the ground state $E_0^{N=3}$ are charge excitation to the $N=2$ and $N=4$ particle sectors with $\Delta E_{\mathrm{ch}} = \frac{1}{2}U + J_{\mathrm{P}} = \frac{1}{2}U - |J_{\mathrm{P}}| \equiv \frac{1}{2}\Delta_{\mathrm{at}}$ as well as spin reconfiguration with $\Delta E_{\mathrm{sp}} = -2J_{\mathrm{P}} = 2|J_{\mathrm{P}}|$ within the $N=3$ charge sector which breaks up orbital singlets and increases the orbital angular momentum from $L=1$ to $L=2$. Thus, the low-energy physics is governed by correlated pair hopping $J_{\mathrm{P}}$ and onsite repulsion $U$. The results presented in Fig.~4 of the main text can be understood from this local limit by addition of the kinetic hopping. In this work, we do not address the regime of charge disproportionation for $\Delta_{\mathrm{at}}<0$ (i.e.~$U<2|J_{\mathrm{P}}|$), where the $N=3$ particle sector is at a higher energy level compared to the $N=2$ and $N=4$ sectors [\href{https://doi.org/10.1103/PhysRevLett.122.186401}{82}].

\begin{table}[t]
	\caption{\textbf{Spectrum of the local Kanamori--Hubbard Hamiltonian for a three-orbital system at half-filling.} The eigenenergies $E_n^N$ are sorted in descending order in each charge sector of particle number $N$ for $J<0$ where the contribution of the correlated pair hopping $J_{\mathrm{P}}$ ($J_{\mathrm{P}}\equiv J$ in our calculations) is explicitly stated. Each state is characterized by total spin $S$, orbital angular momentum $L$, and respective degeneracy $(2S+1)(2L+1)$ with $X = \langle\hat{X}\rangle$ ($X=N,S,L$). The corresponding eigenstates $|\upphi^N_n\rangle$ are given for $N\leq3$ since the $N\geq4$ states can be constructed from particle-hole symmetry. The eigenstates in blue color are those depicted in Fig.~4c of the main text.}
	\label{tab:Local_eigenstates}\vspace{1em}
	\includegraphics[width=1\textwidth]{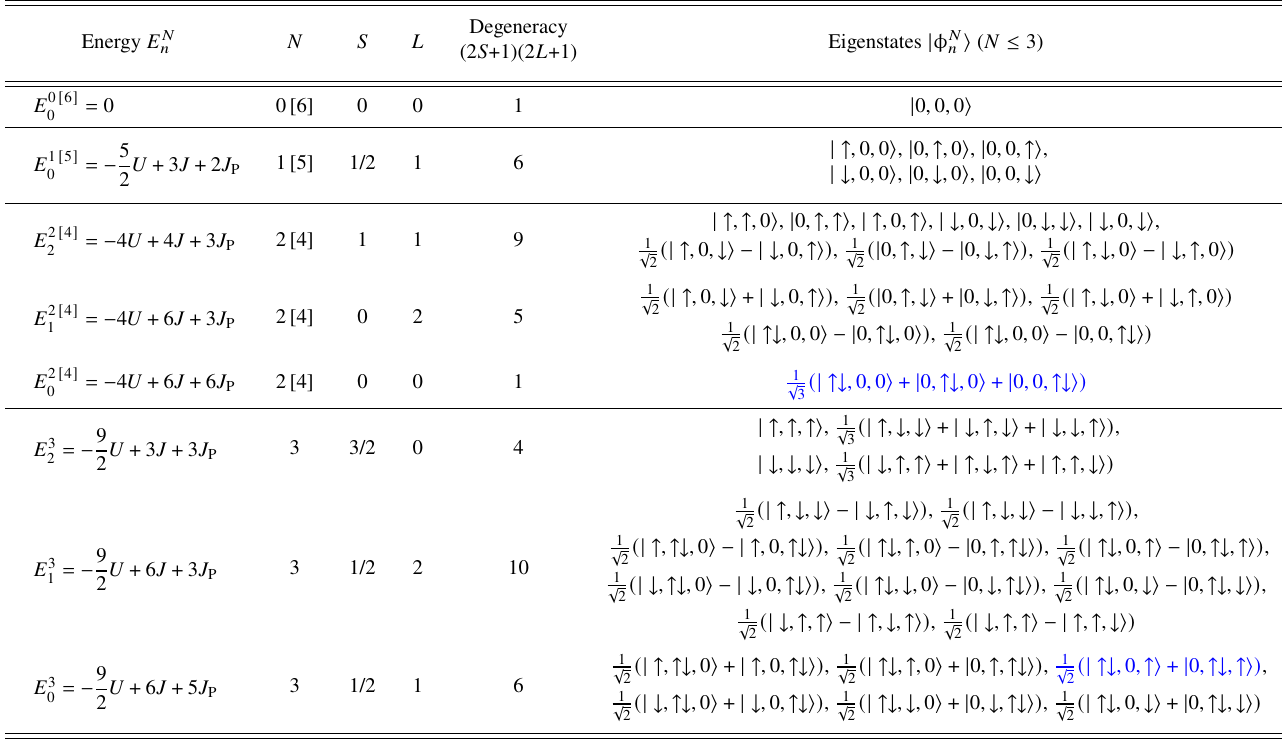}
\end{table}

%
\printbibliography[title=References]